\documentclass[aps,twocolumn,twoside,secnumarabic,balancelastpage,amsmath,amssymb,hyperref=pdftex,superscriptaddress 
]{revtex4}
\usepackage{graphics}      
\usepackage[pdftex]{graphicx}      
\usepackage{longtable}     
\usepackage{bm}            
\usepackage{verbatim}
\usepackage{mathtools}
\usepackage[colorlinks=true]{hyperref}
\usepackage{epstopdf}

\DeclareMathOperator{\Tr}{Tr}

\begin{document}

\title{Evolution of Hall resistivity and spectral function with doping in the SU(2) theory of cuprates}
\date{\today}

\author{C. Morice}
\affiliation{Institut de Physique Th\'eorique, CEA, Universit\'e Paris-Saclay, Saclay, France}
\author{X. Montiel}
\affiliation{Department of Physics, Royal Holloway, University of London, Egham, Surrey, United Kingdom}
\author{C. P\'epin}
\affiliation{Institut de Physique Th\'eorique, CEA, Universit\'e Paris-Saclay, Saclay, France}

\begin{abstract}
Recent transport experiments in the cuprate superconductors linked the opening of the pseudogap to a change in electronic dispersion [S. Badoux et al., Nature 531, 210 (2015)]. Transport measurements showed that the carrier density sharply changes from $x$ to $1+x$ at the pseudogap critical doping, in accordance with the change from Fermi arcs at low doping to a large hole Fermi surface at high doping. The SU(2) theory of cuprates shows that short-range antiferromagnetic correlations cause the arising of both charge and superconducting orders, which are related by an SU(2) symmetry. The fluctuations associated with this symmetry form a pseudogap phase. Here we derive the renormalised electronic propagator under the SU(2) dome, and calculate the spectral functions and transport quantities of the renormalised bands. We show that their evolution with doping matches both spectral and transport measurements.
\end{abstract}

\maketitle

\section{Introduction}

Two of the most striking features of cuprate superconductors are their enigmatic pseudogap phase \cite{Alloul89, Warren89} and how they evolve from a Mott insulator to a correlated metal with doping. Both features have been widely studied during the last thirty years \cite{Norman03, Carlson:2004hn, Lee06, LeHur:2009iw, Rice12, Norman14, Carbotte:2011ip, Eschrig:2006ky, Fradkin:2015ch, Kloss:2016hu} and many scenarios have been proposed to explain the physics of cuprate superconductors, based on antiferromagnetic fluctuations \cite{abanov03, Norman03, sfbook}, strong correlations
\cite{Sorella02, Lee06, Gull:2013hh}, loop currents \cite{Kotliar90, Varma97}, emergent symmetry models \cite{Zhang97, Demler04} and particle-hole patches \cite{Kloss15a, Montiel2017}.

Recent transport experiments at high magnetic field in YBCO \cite{Badoux2015}, Nd-LSCO \cite{Collignon2016} and LSCO \cite{Laliberte2016} showed that these two features are intrinsically linked. Hall coefficient and resistivity measurements indeed yielded a sharp change of the carrier density at the pseudogap critical doping $x^*$, from $x$ at low doping to $1+x$ at high doping \cite{Badoux2015, Collignon2016, Laliberte2016}. Resolving the difference between this critical doping and others, such as the one corresponding to the Fermi surface reconstruction caused by the arising of the charge density wave phase, was made possible by the use of samples with adjacent dopings \cite{Badoux2015, Collignon2016, Laliberte2016, Badoux:2016kg}.

At low doping ($x<x^*$), this change in carrier density is consistent with ARPES experiments which show small Fermi arcs corresponding to $x$ carriers per unit cell \cite{Shen:2005ir}. At high doping ($x>x^*$), the carrier density dependence is in agreement with the quantum oscillation measurements which find a large hole Fermi-surface enclosing a $1+x$ volume, in agreement with band structure calculations \cite{Vignolle2008, Sebastian2011}. It is also reminiscent of earlier transport measurements which detected a peak in the Hall resistivity at low temperature and optimal doping, then attributed to a change in the Fermi surface \cite{Balakirev2003, Balakirev2009}.

Several models have been suggested to explain this change in the carrier density. Some are based on  strong coupling and topological order \cite{Sachdev2016}, which give small Fermi pockets at low temperature which enlarge when temperature rises. The Fermi surface reconstruction caused by charge ordering has also been considered \cite{Zou2017, Sharma2017}. Others are based on long-range fluctuations, either superconducting \cite{Emery1995, Kanigel:2008wm}, antiferromagnetic \cite{Yang2006, Storey2016, Chatterjee2017}, or related to a charge density wave order \cite{Caprara2016} which yield a large Fermi surface gapped at low temperature. Here we discuss a theory related to the latter category, where the fluctuations ensue from the emergence of an SU(2) symmetry between the charge and superconducting operators \cite{Montiel2017}, which both stem from antiferromagnetic correlations.

Emerging symmetry theories have been the subject of controversy, especially since it was argued that the emergence of multiple orders at the pseudogap temperature $T^*$ was due to an ``ineluctable complexity" \cite{Fradkin15}. However, the emergence of an SU(2) symmetry at $T^*$ does account for many features of the phase diagram, in particular for the nematic and loop current responses at $T^*$ \cite{Morice2017skyrmions}.

In this paper, we study electronic transport and spectral functions in the SU(2) theory for cuprate superconductors. We first lay out the main features of the SU(2) theory which is based on short-range antiferromagnetic correlations. We then derive a minimal model for this theory. We come back to the full theory and derive the electron propagator in the pseudogap phase, and finally use it to calculate conductivities, Hall resistivity, and spectral functions. Our results agree with both spectral \cite{Vishik2012} and transport \cite{Badoux2015} measurements.

\section{SU(2) theory}
\label{SU(2) theory}

\subsection{Decoupling the short-range antiferromagnetic Hamiltonian}

Here we outline the salient features of the SU(2) theory of the cuprates, developed in \cite{Montiel2017, Montiel2017sr}. We start by considering a model of short-range antiferromagnetic correlations ($t$-$J$), widely studied in the context of high-temperature superconductivity \cite{Lee06}:
\begin{equation}
H = \sum_{\langle i,j \rangle, \sigma} t_{ij} \psi^\dagger_{i,\sigma} \psi_{j,\sigma} + \frac{1}{2} \sum_{\langle i,j \rangle} J_{ij} \vec{S}_i \cdot \vec{S}_j
\label{eq:t-J Hamiltonian}
\end{equation}
where $i$ and $j$ are lattice sites indices, $\sigma$ and $\sigma'$ are spin indices, $\psi^\dagger$ is the electron creation operator, $\vec{S}_i = \sum_{\sigma,\sigma'} \psi^\dagger_{i\sigma} \vec{\sigma}_{\sigma \sigma'} \psi_{i\sigma'}$ and $\vec{\sigma}$ is the vector of Pauli matrices. This model can be decoupled in the charge and superconducting channels, which yields two corresponding order parameters. First, the charge order parameter:
\begin{equation}
\chi_{k,k'} = \left\langle \sum_\sigma \psi^\dagger_{k,\sigma} \psi_{k',\sigma} \right\rangle
\end{equation}
where $k$ and $k'$ are combined momentum and frequency indices. Here, $\textbf{k}' = \textbf{k} + \textbf{Q}_0$, where $\textbf{Q}_0$ is the charge ordering wave-vector. The second order parameter is the superconducting order parameter:
\begin{equation}
\Delta_{k,k'} = \left\langle \sum_\sigma \sigma \psi^\dagger_{k,\sigma} \psi^\dagger_{k',-\sigma} \right\rangle
\end{equation}
where $\textbf{k}' = -\textbf{k}$. We consider any charge order parameter whose ordering wave-vectors $\textbf{Q}_0$ map hot-spots, which are the point of the Fermi surface separated by the antiferromagnetic ordering wave-vector $(\pi,\pi)$, onto one another.

It was shown that these various charge orders were degenerate with the superconducting order at the hot-spots in a regime of intermediate coupling $J$. This intermediate coupling corresponds to $J$ larger than the energy of the bottom of the conduction band in the anti-nodal region of the first Brillouin zone but smaller than the total bandwidth. The anti-nodal region is the region furthest away from the nodes of the $d$-wave superconducting gap \cite{Montiel2017}.

\subsection{SU(2) fluctuations}

We define the operators:
\begin{align}
\eta^+ = &\sum_\textbf{k} \psi^\dagger_{\textbf{k}, \uparrow} \psi^\dagger_{-\textbf{k}+\textbf{Q}_0,\downarrow}
\\
\eta^- = &\left( \eta^+ \right)^\dagger
\\
\eta^z = &\sum_\textbf{k} \sum_\textbf{k} \psi^\dagger_{\textbf{k}, \uparrow} \psi_{\textbf{k},\uparrow} + \psi^\dagger_{-\textbf{k}+\textbf{Q}_0, \downarrow} \psi_{-\textbf{k}+\textbf{Q}_0,\downarrow} - 1
\end{align}
which form an SU(2) algebra and are thus called SU(2) operators \cite{Montiel2017}. It was shown that, for large values of $J$, the SU(2) operators rotate $\Delta$ and $\chi$ on one another on a line of the Brillouin zone which goes through the hot-spots \cite{Montiel2017}. The two order parameters are therefore related by an exact SU(2) symmetry \cite{Montiel2017}. This generalises some earlier work derived from the spin-fermion model and limited to the hot spots \cite{Efetov13}. The conservation of this SU(2) symmetry leads to fluctuations, whose interaction with the fermions opens a gap in the anti-nodal region of the Brillouin zone \cite{Montiel2017}.

These fluctuations were also shown to raise the degeneracy between the various charge orders, and to select a set of $\textbf{q}$-vectors, including $\textbf{q}=0$ and vectors parallel to the two reciprocal lattice axes \cite{Montiel2017, Montiel2017sr}. This makes this theory intrinsically multi-$\textbf{q}$, i.e.\ with multiple charge ordering wave-vectors.

\begin{figure}
\centering
\includegraphics[width=8cm]{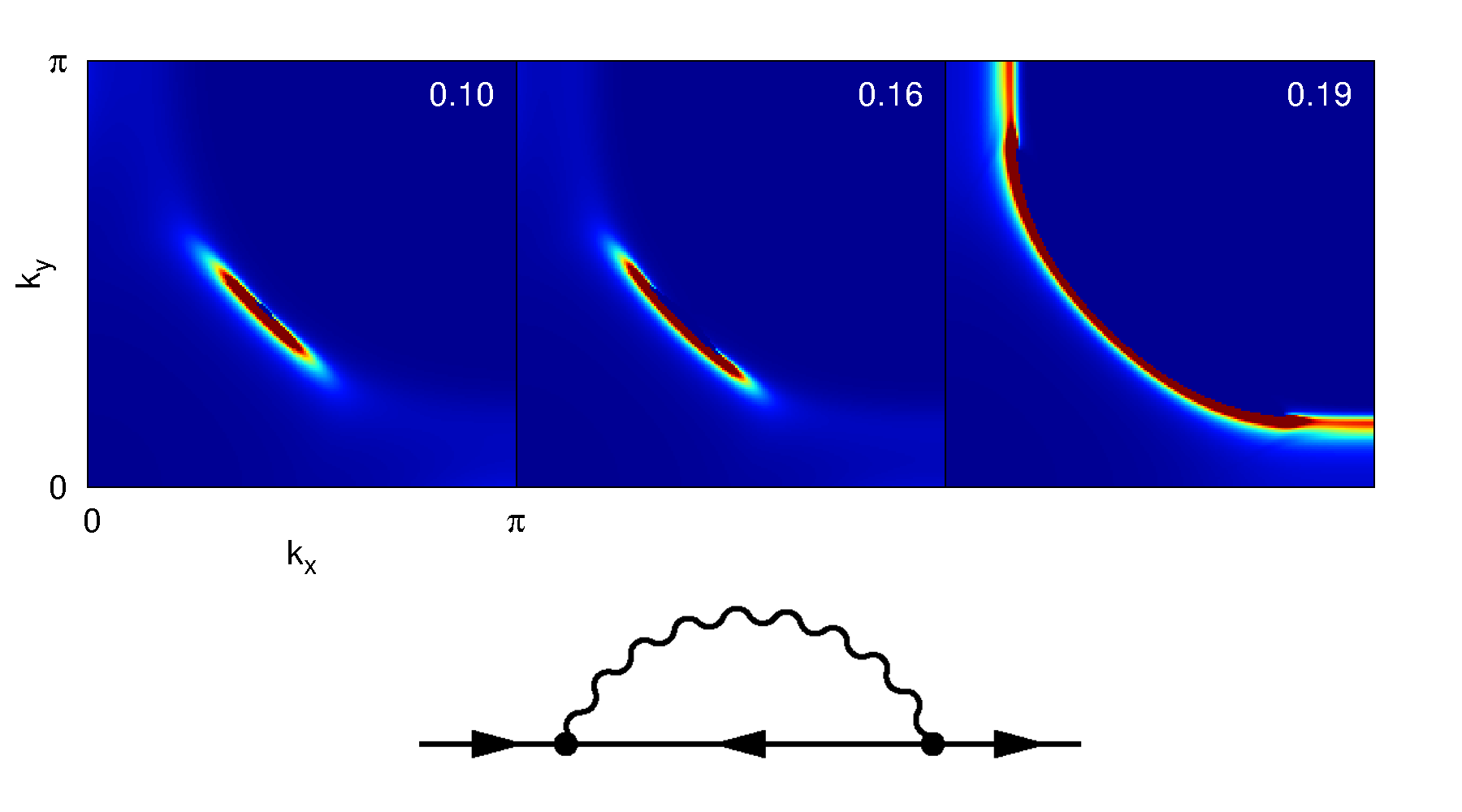}
\caption{Top: evolution of the spectral functions at zero frequency with doping. Note that because the renormalised bands are symmetrical with respect to zero energy, both spectral functions are equal at zero frequency. We therefore only plot one of them. Bottom: self-energy diagram for the renormalisation of the fermionic propagator. The straight lines are bare electron lines, while the wiggly line is the SU(2) fluctuations line.}
\label{fig:spectral}
\end{figure}

This SU(2) theory of the pseudogap has been shown to elucidate many characteristics of the cuprate superconductors, including Raman scattering \cite{Montiel15a}, fixed-doping ARPES \cite{Montiel:2016it} and inelastic neutron scattering \cite{MontielNeutrons2017} responses, as well as the strange metal behaviour \cite{Montiel2017}. Moreover, the recent measurement of pairing fluctuations in the pseudogap region \cite{Hsu2017} is in line with the presence of SU(2) fluctuations \cite{Morice2017skyrmions}. Note that this theory is only based on the presence of antiferromagnetic correlations in the system. In particular, it does not require the presence of any long-range antiferromagnetic fluctuations, nor the existence of an antiferromagnetic quantum critical point.

The composite order parameter corresponding to this SU(2) order is a $2 \times 2$ matrix:
$$
b
\left(
\begin{matrix}
\chi       & \Delta \\
-\Delta^\dagger & \chi^\dagger
\end{matrix}
\right)
\text{ where }
|\chi|^2 + |\Delta|^2 = 1
$$
with $\chi$ the charge order parameter, $\Delta$ the superconducting order parameter and $b$ the SU(2) phase \cite{Montiel2017}. Note that this order parameter is intrinsically not abelian. It is SU(2)-symmetric, meaning that there exists a set of operators forming an SU(2) algebra under which this composite order parameter is invariant \cite{Montiel2017}.

\section{Minimal model}

Here we consider a simpler version of the SU(2) theory, which was studied previously in the context of ARPES \cite{Montiel15a}, Raman \cite{Montiel:2016it} and inelastic neutron scattering responses \cite{MontielNeutrons2017}.

The minimal model for the SU(2) theory is obtained by performing a mean-field decoupling of the $t$-$J$ model in two channels \cite{Montiel15a, MontielNeutrons2017}. The first channel is, like for the general case, the $d$-wave superconducting channel $\Delta_\textbf{k}$. The second channel is the Resonant Excitonic State (RES) channel, which is a specific multi-$\textbf{q}$ charge order. It corresponds to $\chi_\textbf{k}$ with a $\textbf{k}$-dependent ordering wave vector $2 p_F (\textbf{k})$. It maps one side of the Fermi surface on the other \cite{Montiel15a, MontielNeutrons2017}. More precisely, for $\textbf{k}$ on the Fermi surface, $2 p_F (\textbf{k}) = - \textbf{k}$. The RES corresponds to patches of charge order arranged in real space \cite{Montiel15a, MontielNeutrons2017, Montiel2017sr}.

\begin{widetext}
We start from the $t$-$J$ model (equation \eqref{eq:t-J Hamiltonian}) and Fourier transform the fermionic operators:
\begin{align}
H = &\sum_{\textbf{k}, \sigma} \xi_\textbf{k} \psi^\dagger_{\textbf{k}, \sigma} \psi_{\textbf{k}, \sigma} + \frac{1}{2} \sum_{\textbf{k}, \textbf{k}', \textbf{q}} \sum_{\alpha, \beta, \gamma, \delta} J_{\textbf{q}} \psi^\dagger_{\textbf{k}, \alpha} \vec{\sigma}_{\alpha \beta} \psi_{\textbf{k}+\textbf{q}, \beta} \cdot \psi^\dagger_{\textbf{k}'+\textbf{q}', \gamma} \vec{\sigma}_{\gamma \delta} \psi_{\textbf{k}', \delta}
\label{eq:t-J Fourier}
\end{align}
where $\alpha$, $\beta$, $\gamma$ and $\delta$ are spin indices, and $\xi_{\textbf{k}}$ is the free electron dispersion. The mean-field decoupling of the Hamiltonian \eqref{eq:t-J Fourier} in the superconducting and RES channels yields the effective Hamiltonian :
\begin{equation}
H_{eff} = - \sum_{\textbf{k}, \sigma} \Psi^\dagger_{\textbf{k}, \sigma} G^{-1}_{\text{min}} \Psi_{\textbf{k}, \sigma}
\end{equation}
where $\Psi^\dagger_{\textbf{k} \sigma}$ is the four-states spinor $\left( \psi^\dagger_{\textbf{k}, \sigma}, \psi_{-\textbf{k} - 2p_F(-\textbf{k}), -\sigma}, \psi^\dagger_{\textbf{k}- 2p_F(\textbf{k}) \sigma}, \psi_{-\textbf{k}, -\sigma} \right)$
and:
\begin{equation}
G^{-1}_{\text{min}}= 
\begin{pmatrix}
\omega - \xi_\textbf{k} & 0 & \chi_\textbf{k} & \Delta_\textbf{k}\\
0 & \omega + \xi_{-\textbf{k}-2p_F(-\textbf{k})} & \Delta^\dagger_{\textbf{k}-2p_F(\textbf{k})} & -\chi_\textbf{k} \\
\chi^\dagger_\textbf{k} & \Delta_{\textbf{k}-2p_F(\textbf{k})} & \omega - \xi_{\textbf{k}-2p_F(\textbf{k})} & 0 \\
\Delta^\dagger_\textbf{k} & -\chi^\dagger_\textbf{k} & 0 & \omega + \xi_{-\textbf{k}}
\end{pmatrix}
\label{eq:G inverse}
\end{equation}
\end{widetext}
The propagator of the minimal model can be obtained by inverting the matrix \eqref{eq:G inverse} as done in \cite{Montiel:2016it, MontielNeutrons2017}. Close to the Fermi surface, one can approximate $\xi_{\textbf{k}-2p_F(\textbf{k})}$ by $-\xi_{\textbf{k}}$ and $\Delta^\dagger_{\textbf{k}-2p_F(\textbf{k})}$ by $\Delta^\dagger_{\textbf{k}}$ \cite{Montiel:2016it, MontielNeutrons2017}. The renormalised propagator of the minimal model then writes:
\begin{equation}
G_{\text{min}_{11}}(\textbf{k},\omega) = \frac{1}{\omega - \xi_\textbf{k} - \frac{|\chi_\textbf{k}|^2 + |\Delta_\textbf{k}|^2}{\omega + \xi_{\textbf{k}}}}
\label{eq:propagator-minimal}
\end{equation}
where $|\chi_\textbf{k}|^2 + |\Delta_\textbf{k}|^2$ is the SU(2) order parameter which is maximal when the SU(2) symmetry is conserved and $11$ indicates the first-line first-column coefficient of the inverse matrix.

This minimal model makes clear that the gapping of the anti-nodal part of the Brillouin zone is due to the formation of fluctuating pairs: particle-particle pairs for the superconducting order and particle-hole pairs for the charge order, and the fluctuation between these two types of pairs. Antiferromagnetic correlations therefore cause the arising of fluctuating pairs which ``schizophrenically" fluctuate between the particle-particle and the particle-hole channels.

\section{Derivation of the self-energy}

We now come back to the full SU(2) theory outlined in section \ref{SU(2) theory}, and derive the renormalised propagator in this case. Expanding the action of the model of short-range antiferromagnetic correlations linearly for small SU(2) fluctuations, one obtains an effective non-linear $\sigma$-model \cite{Montiel2017}. The effective action for the electrons is obtained by integrating out the SU(2) fluctuations in this model \cite{Montiel2017}: 
\begin{align}
S_{fin} = &-\frac{1}{2} \Tr \sum_{k,k',q,q',\sigma,\sigma'} \sigma \sigma' \left\langle \Delta_{kq}^\dagger \Delta_{k'q'} \right\rangle_Q \nonumber
\\
&\times \psi^\dagger_{k+q,\sigma} \psi^\dagger_{-k+q,\bar{\sigma}} \psi_{-k'+q',\bar{\sigma}'} \psi_{k'+q',\sigma'}
\label{eq:action}
\end{align}
where $k$, $k'$, $q$ and $q'$ are combined momentum and frequency indices, $\sigma$ and $\sigma'$ are spin indices and $\psi^\dagger$ is the electron creation operator. One can simplify \cite{Montiel2017}:
\begin{equation}
\left\langle \Delta_{kq}^\dagger \Delta_{k'q'} \right\rangle_Q = \delta_{\textbf{q},\textbf{q}'} \pi^s_{kk'q}
\end{equation}
where $\pi^s_{kk'q}$ is the SU(2) fluctuations propagator:
\begin{equation}
\pi^s_{kk'q} = M_{0,k} M_{0,k'} \frac{\pi_0}{J_0 \epsilon^2 + J_1(\textbf{v}\cdot \textbf{q})^2 - a_0}
\end{equation}
where $q=(\textbf{q},\epsilon)$, $\pi_0$, $J_0$ and $J_1$ are coefficients, $\textbf{v}$ is the Fermi velocity, $a_0$ is a mass term, and $M_{0,k}$ is the SU(2) form factor, which depends on the position in the Brillouin zone. $M_{0,k}=1$ when the SU(2) symmetry is preserved, and is zero when it is broken. $a_0<0$ when a magnetic field is applied. Inputting this in equation \eqref{eq:action} gives the expression for the self energy:
\begin{equation}
-\Sigma(k) = \frac{1}{2} \sum_{q,\sigma} \pi^s_{kkq} G^0_{-k+q,\bar{\sigma}}
\label{eq:self}
\end{equation}
where $G^0$ is the free electron propagator. This self-energy corresponds to the diagram in figure \ref{fig:spectral}. Approximating the sum yields (see Supplemental Material for details):
\begin{equation}
\Sigma(\textbf{k},\omega) = B \frac{M_{0,k}^2}{i\omega + \xi_{\textbf{k}}}
\end{equation}
where $B$ is a parameter, $k=(\textbf{k},\omega)$ and $\xi_{\textbf{k}}$ is the free electron dispersion. The renormalised electronic propagator therefore is:
\begin{equation}
G(\textbf{k},\omega) = \frac{1}{\omega-\xi_\textbf{k} - B \frac{M_{0,k}^2}{\omega + \xi_{\textbf{k}}}}
\label{eq:propagator}
\end{equation}
We therefore obtain the same propagator as in the minimal model, except that the SU(2) order parameter is replaced by a form factor inherited from the full SU(2) theory, and that we did not need to make any approximation on the dispersion.

\begin{figure}
\centering
\includegraphics[width=8cm]{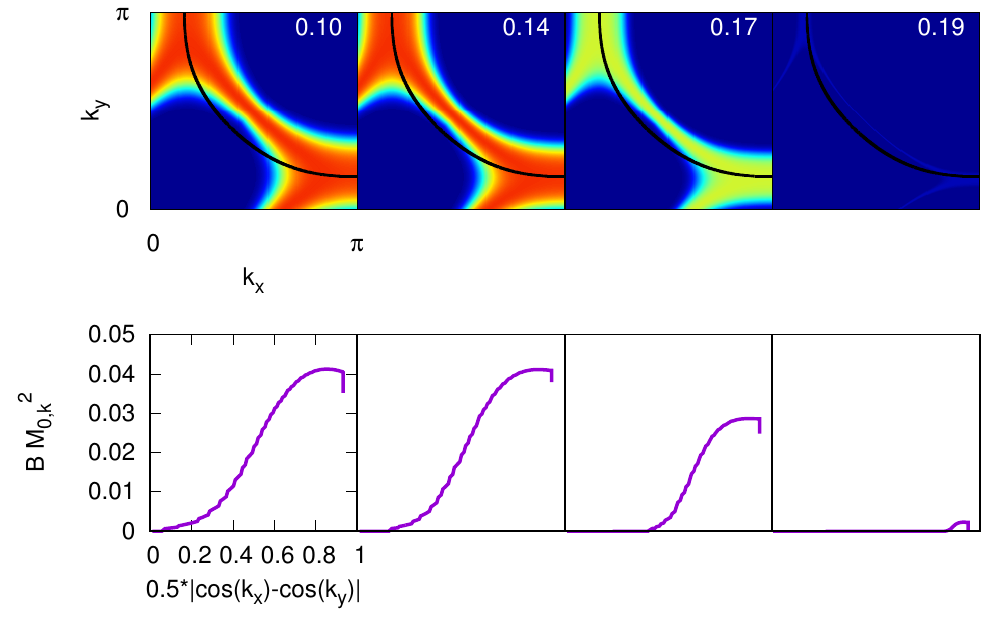}
\caption{Top: evolution of the gap $B M_{0,k}^2$ with doping in a quarter of the Brillouin zone. The color blue is for zero, and the black line represents the Fermi surface. Bottom: gap on the Fermi surface with respect to the $d$-wave factor. A pure $d$-wave gap would be strictly linear.}
\label{fig:gap}	
\end{figure}

\section{Spectral and transport responses}

Separating the renormalised propagator obtained with the full SU(2) theory (equation \eqref{eq:propagator}) in simple elements gives us the expressions for the renormalised bands:
\begin{equation}
E^\pm(\textbf{k}) = \pm \sqrt{ \xi (\textbf{k}, x)^2 + B M_{0,k}^2}
\end{equation}
and corresponding spectral functions:
\begin{equation}
A^\pm (\textbf{k}, \omega) = \frac{1}{\pi} \frac{W^\pm(\textbf{k}) \Gamma^\pm (\textbf{k})}{ \left(\omega - E^\pm(\textbf{k}, x)\right)^2 + \Gamma^\pm (\textbf{k})^2}
\label{eq:spectral-functions}
\end{equation}
where the spectral weights are given by:
\begin{equation}
W^\pm(\textbf{k}) = \frac{1}{2} \left( 1 \pm \frac{\xi (\textbf{k}, x)}{\sqrt{  \xi (\textbf{k}, x)^2 + B M_{0,k}^2}} \right)
\end{equation}
and $\Gamma^\pm$ is the scattering rate of each renormalised band. We follow previous works \cite{Voruganti1992, Storey2016, Eberlein2016} and neglect the scattering between the two renormalised bands. The longitudinal and transverse conductivities are given by \cite{Voruganti1992}:
\begin{equation}
\sigma^\pm_{xx} = -\frac{2 \pi e^2}{V N} \sum_\textbf{k} \left(v_x^\pm (\textbf{k}) \right)^2 \int d\omega \frac{\partial f(\omega)}{\partial \omega} A^\pm(\textbf{k}, \omega)^2
\end{equation}
\begin{align}
\sigma^\pm_{xy} = &\frac{4 \pi^2 e^3}{3 V N} \sum_\textbf{k} v_x^\pm (\textbf{k}) \left( v_x^\pm (\textbf{k}) \frac{\partial v_y^\pm (\textbf{k})}{\partial k_y} -  v_y^\pm (\textbf{k}) \frac{\partial v_y^\pm (\textbf{k})}{\partial k_x} \right) \nonumber
\\
&\times \int d\omega \frac{\partial f(\omega)}{\partial \omega} A^\pm(\textbf{k}, \omega)^3
\end{align}
The integral over frequency can be simplified using the standard approximation \cite{Mahan}:
\begin{equation}
\int_{-\infty}^\infty d\omega \left( \frac{\Gamma^\pm(\textbf{k})}{\left( \omega - E^\pm(\textbf{k}) \right)^2 + \Gamma^\pm(\textbf{k})^2} \right)^2 = \frac{\pi}{2} \frac{1}{\Gamma^\pm(\textbf{k})}
\end{equation}
which gives:
\begin{equation}
\sigma^\pm_{xx} = \frac{e^2}{V N} \sum_\textbf{k} \left(v_x^\pm (\textbf{k}) \right)^2 \frac{W^\pm(\textbf{k})^2}{\Gamma^\pm(\textbf{k})} \frac{\beta e^{\beta E^\pm(\textbf{k})}}{\left( e^{\beta E^\pm(\textbf{k})} + 1 \right)^2}
\label{eq:sigma_xx approx}
\end{equation}
We generalise this approach to the cubic case and obtain:
\begin{align}
\sigma^\pm_{xy} = &-\frac{e^3}{2 V N} \sum_\textbf{k} v_x^\pm (\textbf{k}) \left( v_x^\pm (\textbf{k}) \frac{\partial v_y^\pm (\textbf{k})}{\partial k_y} -  v_y^\pm (\textbf{k}) \frac{\partial v_y^\pm (\textbf{k})}{\partial k_x} \right) \nonumber
\\
& \times \frac{W^\pm(\textbf{k})^3}{\Gamma^\pm(\textbf{k})^2} \frac{\beta e^{\beta E^\pm(\textbf{k})}}{\left( e^{\beta E^\pm(\textbf{k})} + 1 \right)^2}
\label{eq:sigma_xy approx}
\end{align}
These expressions allow us to calculate the Hall resistance \cite{Voruganti1992}:
\begin{equation}
R_H = \frac{\sigma^+_{xy} + \sigma^-_{xy}}{\left( \sigma^+_{xx} + \sigma^-_{xx} \right)^2}
\end{equation}

The magnitude of the gap depends on the conservation of the SU(2) symmetry. Indeed, when the SU(2) symmetry is conserved, the SU(2) fluctuations are maximal, and therefore the gap is fully open. When the SU(2) symmetry is sufficiently broken, there are no SU(2) fluctuations and the gap is closed. In order to quantify how much the SU(2) symmetry is broken, we define $\Delta \xi_\textbf{k} = \frac{1}{2} \left( \xi_\textbf{k} + \xi_\textbf{k+Q} \right)$ which we name the SU(2) symmetry-breaking dispersion, or SU(2) line. If $\Delta \xi_\textbf{k}$ is zero, the SU(2) symmetry is conserved, hence we must have $M_0^2 = 1$ to open the gap. Conversely, if the SU(2) symmetry is broken, $\Delta \xi_\textbf{k}$ is large, and we must have $M_0^2 = 0$. In order to interpolate between these two points, we parametrize the symmetry breaking coefficient in the free energy using a smooth step function:
\begin{equation}
M_0^2 = \frac{1}{e^{ 30*\left(\frac{\Delta \xi_\textbf{k}}{\Delta_{SU2}} \right)^2 - 0.02} + 1}
\end{equation}
where $\Delta_{SU2}$ is the magnitude of the SU(2) gap. The SU(2) wave vector $Q_0$ is chosen as the vector between the two closest hot-spots, following previous studies \cite{Montiel2017}. We set $M_0$ to zero when smaller than one hundredth. The $\Delta_{SU2}$ parameter represents the magnitude of the pseudogap order parameter in the SU(2) theory and was parametrised by:
\begin{equation}
\Delta_{SU2} = \left( \frac{1}{e^{(x - 0.175) \times 170} + 1} - 0.018 \right) \times 0.58
\end{equation}
For consistency we also set $B=\Delta_{SU2}$. We use $\Gamma^\pm = 0.01 \times t_0$, and set $x^*=0.2$. We use the electronic dispersion used in a previous work \cite{Storey2016}, and shown to properly replicate the doping dependence of the Hall number for $x>x^*$.

\begin{figure}
\centering
\includegraphics[width=8cm]{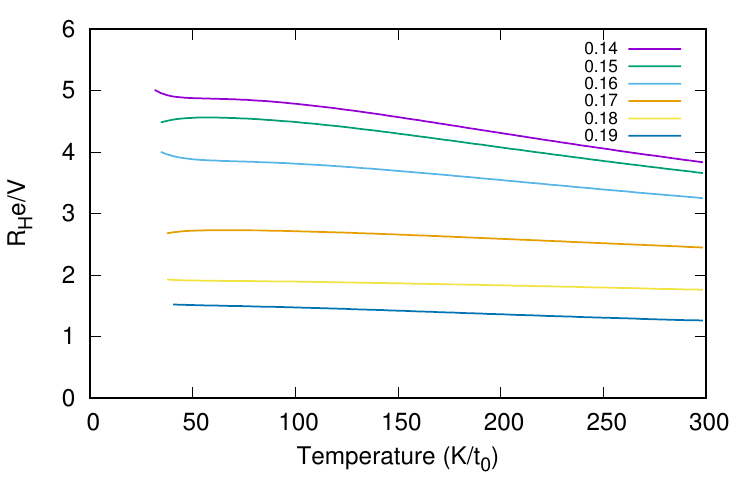}
\caption{Hall resistance in volume units with respect to temperature in Kelvin per units of $t_0$, for a range of hole dopings.}
\label{fig:RH}
\end{figure}

\section{Results}

We calculated the magnitude of the gap $B \times M_0^2$ over the Brillouin zone and on the Fermi surface (Figure \ref{fig:gap}). The gap opens along the SU(2) line, as found previously \cite{Montiel2017}. The SU(2) line crosses the Fermi surface at the hot spots, consequently of our choice of ordering wave-vector. The gap opens in the antinodal zone and is closed in the nodal zone. It gets both thinner and smaller in magnitude with rising doping and finally vanishes at the critical doping. This can be compared with ARPES data which showed that the pseudogap was closed in the nodal zone \cite{Vishik2012}. Our data fits qualitatively these experimental results, unlike methods based on a pure $d$-wave gap (i.e.\ a gap linear in the $d$-wave factor).

\begin{figure}
\centering
\includegraphics[width=8cm]{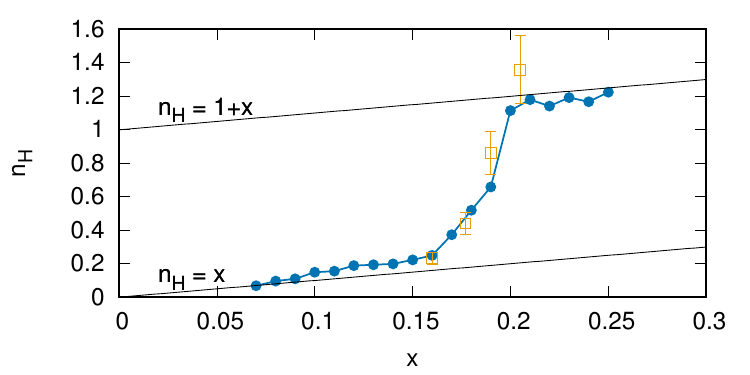}
\caption{Hall number with respect to doping (blue). The experimental values from \cite{Badoux2015} (orange) and the low-doping and high-doping asymptotes (black) are also plotted for reference. Note that $x^*=0.2$.}
\label{fig:nH}
\end{figure}

The spectral functions of the two renormalised bands at zero frequency were calculated using equation \eqref{eq:spectral-functions} (Figure \ref{fig:spectral}). The dispersion in the denominator of the self-energy (equation \eqref{eq:self}), which corresponds to the dispersion of the SU(2) fluctuations bosonic mode, is the opposite of the bare electronic dispersion. The self-energy therefore diverges on the Fermi surface. This also means that the renormalised bands are equal to each other up to a sign, and therefore the spectral functions at zero frequency are equal. Consequently, we only plot one of them. The absence of gap in the nodal region causes the formation of a Fermi arc around the nodal point. This Fermi arc gets larger with larger doping, until it forms the whole Fermi surface at the critical doping $x^*$. The gap remains open at the hot spots for $x<x^*$, and closes near the Brillouin zone edge slightly before (see Supplemental Material).

Using equations \eqref{eq:sigma_xx approx} and \eqref{eq:sigma_xy approx}, we calculated the evolution of the Hall resistivity with temperature (Figure \ref{fig:RH}). For completeness, we also plotted the evolution of the longitudinal and transverse conductivities (see Supplemental Material). For each doping, the Hall resistivity rises with decreasing temperature and saturates at low temperature. This rise is lower for higher dopings, and almost absent close to the critical doping. Note that the calculation does not reach absolute zero. This is due to the exponentials in the expressions for conductivities growing larger than the computational maximum. The zero-temperature Hall number $n_H=V/eR_H$ (Figure \ref{fig:nH}) sharply changes from $x$ to $1+x$ close to the critical doping, in agreement with experimental measurements on YBCO \cite{Badoux2015} and Nd-LSCO \cite{Collignon2016}. The $1+x$ evolution of $n_H$ at high doping corresponds to the standard carrier density for a hole pocket, and is similar to the one obtained in other theories \cite{Storey2016}.

We compared these results to using a pure $d$-wave gap such as the one used in previous studies \cite{Storey2016}. Naturally, because the dispersion of the bosonic mode is equal to minus the electronic dispersion, the gap opens everywhere but at the nodal points. Therefore the Hall resistance diverges at low temperature (see Supplemental Material). However one can measure the evolution of finite temperature values of the Hall resistance with doping. Interestingly, this displays a transition from $x$ to $1+x$, in fine agreement with experiments (see Supplemental Material).

\section{Discussion}

Our calculation of the evolution of the gap on the Fermi surface with doping closely resembles ARPES data \cite{Vishik2012} (Figure \ref{fig:spectral}). Indeed, measurement of the photoemission gap above $T_c$ found that a $d$-wave dependence could not describe what was seen experimentally: the gap is measured to be zero for a segment of the $d$-wave factor which grows with doping, unlike a $d$-wave gap which would be linear here. Experimental data then finds a close-to-linear increase in the gap, followed by a saturation at intermediate dopings \cite{Vishik2012}. Our data satisfactorily fits experimental data at low and intermediate doping. Indeed the saturation is not observed in the sample with the lowest doping, but data points for a $d$-wave factor larger than $0.8$ have larger error bars and could fit a saturation, knowing that this is precisely the $d$-wave factor where we find a saturation. At high doping however, we find a segment for which the gap is zero which is much larger that in experiments. But this part of the experimental data is more noisy and closer to zero which makes us think that this could be due to experimental difficulties. Indeed, this observation is in contradiction with the length of the Fermi arc measured with ARPES, which can be seen to be large in other studies when plotted on the Brillouin zone \cite{Shen:2005ir}, while the pseudogap is finite for a $d$-wave factor as low as 0.5. Finally, the size of the gap at its maximum was measured to be about 50 meV at low doping, 40 meV at intermediate doping and 20 meV a high doping \cite{Vishik2012}. Our calculations reproduce this trend, although the sharp drop of this maximum at high doping has yet to be compared with experiments very close to the critical doping. We therefore conclude that our calculations are in agreement with $d$-wave-factor resolved ARPES data \cite{Vishik2012}.

Our calculations therefore yield the arising of Fermi arcs, not Fermi pockets, in the pseudogap phase. This seems to be in contradiction with the Luttinger sum rule, which states that the volume of the Fermi surface is equal to the number of carriers. Indeed, here in the pseudogap phase this volume is ill defined, since parts of its boundary have vanished. However, in our case, it is the fluctuations between superconductivity and charge order which gap the Fermi surface. We extrapolate this theory at zero temperature, in order to replicate the experimental procedure. But the pseudogap is not a zero-temperature ground state in this theory, and therefore there is no breaking of the Luttinger theorem. At zero temperature, one would obtain either superconductivity or charge order, possibly with defects such as superconducting filaments in the charge ordered phase \cite{Morice2017skyrmions, Hsu2017}. The fact that we obtain arcs and not pockets is similar to the case of a multi-$\textbf{q}$ charge order, such as the one considered in the minimal model \cite{Montiel:2016it, Montiel15a, MontielNeutrons2017}, or to the case of a superconducting gap . Indeed, there, the perfect nesting of the Fermi surface means that there is no rise of electron or hole pockets, nor any reconfiguration of the Fermi surface. Instead, the Fermi surface is gapped in the anti-nodal region.

The Hall number is a measurement of the number of carriers, and accordingly we find that the length of the Fermi arcs has a similar evolution to it. This also means that in order for the Hall number to reach the $1+x$ line, the gap has to close near the edge of the Brillouin zone, in order to allow the formation of a second small Fermi arc per quarter of the Brillouin zone, close to the zone edge, separated from the first one by the two hot spots. Here, this extra Fermi arc appears at dopings higher than $0.19$ (see Supplemental Material).

The evolution of the Hall number depends on the parametrisation of the gap, as in any phenomenological model. The specific choices we made here correspond to the measurements on YBCO \cite{Badoux2015}. However the fact that this system goes from small Fermi arcs to a large hole pocket does not rely on fitting. Only the width of the transition can be tuned. We did not use, unlike many other parametrisation of the pseudogap, a linear dependence with respect to doping \cite{Storey2016}. This linear dependence does not fit experimental data, even in the studies that use it. Indeed, the YRZ model with the published linear pseudogap gives $n_H$ under the $x$ line for $x<0.10$ (see Supplemental Material) \cite{Storey2016}. This does not challenge the ability of the YRZ model to replicate experimental data, but only stresses that, there too, a linear dependence of the pseudogap in doping is inadequate. Note that the evolution of $n_H$ at high doping, which follows the $1+x$ line, is the same as the one in the YRZ model since at that point the gap is entirely closed and therefore the theories no longer differ.

The fact that the Hall number goes as $x$ at low doping is a matter of fitting in both the YRZ model and the current study. However it is also a physical consequence of the fact that both models are based on antiferromagnetic correlations. Doping such a system will cause the arising of Fermi pockets of size $x$, which will then either grow with doping in the case of the YRZ model, or give birth to arcs in the case of the SU(2) theory, at the doping at which the SU(2) fluctuations start dominating the antiferromagnetic fluctuations.

The choice of ordering vector $Q_0$ at the hot spots has been made according to previous studies \cite{Montiel2017}. We did however replicate the calculation for two other $Q_0$ ordering vectors, one taken at the Brillouin zone edge, and one linking two diagonally placed hot spots. The first choice does not impact the calculations much, but the second produces an early transition of $n_H$, which reaches the $1+x$ line at $x=0.19$ (see Supplemental Material). What is crucial for the transition of the Hall number is therefore not the precise choice of $Q_0$, but that the anti-nodal region is implicated.

\section{Conclusion}

Here we derived the renormalised electronic propagator in the pseudogap phase, in the framework of the SU(2) theory. Comparing it with the one obtained for a minimal model of fluctuating pairs tells us that such a simple theory is compatible with experimental observations. This is true as long as the pairing energy comes from short-range antiferromagnetic correlations, which are key to make the link with the low doping part of the phase diagram. Finally, the choice of charge ordering wave vector does not have an impact on the result, as long as it involves the anti-nodal region.

It is striking that our results fit experimental results corresponding both to transport and spectral probes closely. Indeed they agree to a remarkable extent with two types of ARPES measurements: over the Brillouin zone and resolved with respect to the $d$-wave factor (Figures \ref{fig:spectral} and \ref{fig:gap}, respectively). They also quantitatively reproduce the evolution of the Hall number with doping (Figure \ref{fig:nH}). These results are directly inferred from the SU(2) theory of superconductors, which is an effective theory derived directly from a model of antiferromagnetism with short-range coupling. Moreover, the SU(2) theory has been shown to agree well with other experimental signatures, such as details in energy-resolved ARPES spectra \cite{Montiel:2016it} and in Raman scattering \cite{Montiel15a} and neutrons \cite{MontielNeutrons2017} experiments. The agreement with such a wide range of experiments is indeed encouraging.

\section*{Acknowledgments}

We would like to thank Sven Badoux, Nicolas Doiron-Leyraud, Cyril Proust, Suchitra Sebastian, Yvan Sidis, Louis Taillefer, Andr\'e-Marie Tremblay and Simon Verret for stimulating discussions. This work received financial support from the ANR project UNESCOS ANR-14-CE05-0007 and the ERC, under grant agreement AdG-694651-CHAMPAGNE. The authors also would like to thank the IIP (Natal, Brazil), where parts of this work were done, for hospitality. 

\bibliographystyle{apsrev4-1}
\bibliography{Cuprates}

\begin{thebibliography}{53}%
\makeatletter
\providecommand \@ifxundefined [1]{%
 \@ifx{#1\undefined}
}%
\providecommand \@ifnum [1]{%
 \ifnum #1\expandafter \@firstoftwo
 \else \expandafter \@secondoftwo
 \fi
}%
\providecommand \@ifx [1]{%
 \ifx #1\expandafter \@firstoftwo
 \else \expandafter \@secondoftwo
 \fi
}%
\providecommand \natexlab [1]{#1}%
\providecommand \enquote  [1]{``#1''}%
\providecommand \bibnamefont  [1]{#1}%
\providecommand \bibfnamefont [1]{#1}%
\providecommand \citenamefont [1]{#1}%
\providecommand \href@noop [0]{\@secondoftwo}%
\providecommand \href [0]{\begingroup \@sanitize@url \@href}%
\providecommand \@href[1]{\@@startlink{#1}\@@href}%
\providecommand \@@href[1]{\endgroup#1\@@endlink}%
\providecommand \@sanitize@url [0]{\catcode `\\12\catcode `\$12\catcode
  `\&12\catcode `\#12\catcode `\^12\catcode `\_12\catcode `\%12\relax}%
\providecommand \@@startlink[1]{}%
\providecommand \@@endlink[0]{}%
\providecommand \url  [0]{\begingroup\@sanitize@url \@url }%
\providecommand \@url [1]{\endgroup\@href {#1}{\urlprefix }}%
\providecommand \urlprefix  [0]{URL }%
\providecommand \Eprint [0]{\href }%
\providecommand \doibase [0]{http://dx.doi.org/}%
\providecommand \selectlanguage [0]{\@gobble}%
\providecommand \bibinfo  [0]{\@secondoftwo}%
\providecommand \bibfield  [0]{\@secondoftwo}%
\providecommand \translation [1]{[#1]}%
\providecommand \BibitemOpen [0]{}%
\providecommand \bibitemStop [0]{}%
\providecommand \bibitemNoStop [0]{.\EOS\space}%
\providecommand \EOS [0]{\spacefactor3000\relax}%
\providecommand \BibitemShut  [1]{\csname bibitem#1\endcsname}%
\let\auto@bib@innerbib\@empty
\bibitem [{\citenamefont {Alloul}\ \emph {et~al.}(1989)\citenamefont {Alloul},
  \citenamefont {Ohno},\ and\ \citenamefont {Mendels}}]{Alloul89}%
  \BibitemOpen
  \bibfield  {author} {\bibinfo {author} {\bibfnamefont {H.}~\bibnamefont
  {Alloul}}, \bibinfo {author} {\bibfnamefont {T.}~\bibnamefont {Ohno}}, \ and\
  \bibinfo {author} {\bibfnamefont {P.}~\bibnamefont {Mendels}},\ }\href
  {\doibase 10.1103/PhysRevLett.63.1700} {\bibfield  {journal} {\bibinfo
  {journal} {Phys. Rev. Lett.}\ }\textbf {\bibinfo {volume} {63}},\ \bibinfo
  {pages} {1700} (\bibinfo {year} {1989})}\BibitemShut {NoStop}%
\bibitem [{\citenamefont {Warren}\ \emph {et~al.}(1989)\citenamefont {Warren},
  \citenamefont {Walstedt}, \citenamefont {Brennert}, \citenamefont {Cava},
  \citenamefont {Tycko}, \citenamefont {Bell},\ and\ \citenamefont
  {Dabbagh}}]{Warren89}%
  \BibitemOpen
  \bibfield  {author} {\bibinfo {author} {\bibfnamefont {W.~W.}\ \bibnamefont
  {Warren}}, \bibinfo {author} {\bibfnamefont {R.~E.}\ \bibnamefont
  {Walstedt}}, \bibinfo {author} {\bibfnamefont {G.~F.}\ \bibnamefont
  {Brennert}}, \bibinfo {author} {\bibfnamefont {R.~J.}\ \bibnamefont {Cava}},
  \bibinfo {author} {\bibfnamefont {R.}~\bibnamefont {Tycko}}, \bibinfo
  {author} {\bibfnamefont {R.~F.}\ \bibnamefont {Bell}}, \ and\ \bibinfo
  {author} {\bibfnamefont {G.}~\bibnamefont {Dabbagh}},\ }\href {\doibase
  10.1103/PhysRevLett.62.1193} {\bibfield  {journal} {\bibinfo  {journal}
  {Phys. Rev. Lett.}\ }\textbf {\bibinfo {volume} {62}},\ \bibinfo {pages}
  {1193} (\bibinfo {year} {1989})}\BibitemShut {NoStop}%
\bibitem [{\citenamefont {Norman}\ and\ \citenamefont
  {P\'epin}(2003)}]{Norman03}%
  \BibitemOpen
  \bibfield  {author} {\bibinfo {author} {\bibfnamefont {M.~R.}\ \bibnamefont
  {Norman}}\ and\ \bibinfo {author} {\bibfnamefont {C.}~\bibnamefont
  {P\'epin}},\ }\href {http://stacks.iop.org/0034-4885/66/i=10/a=R01}
  {\bibfield  {journal} {\bibinfo  {journal} {Rep. Prog. Phys.}\ }\textbf
  {\bibinfo {volume} {66}},\ \bibinfo {pages} {1547} (\bibinfo {year}
  {2003})}\BibitemShut {NoStop}%
\bibitem [{\citenamefont {Carlson}\ \emph {et~al.}(2004)\citenamefont
  {Carlson}, \citenamefont {Kivelson}, \citenamefont {Orgad},\ and\
  \citenamefont {{Emery, V. J.}}}]{Carlson:2004hn}%
  \BibitemOpen
  \bibfield  {author} {\bibinfo {author} {\bibfnamefont {E.~W.}\ \bibnamefont
  {Carlson}}, \bibinfo {author} {\bibfnamefont {S.~A.}\ \bibnamefont
  {Kivelson}}, \bibinfo {author} {\bibfnamefont {D.}~\bibnamefont {Orgad}}, \
  and\ \bibinfo {author} {\bibnamefont {{Emery, V. J.}}},\ }\href {\doibase
  10.1007/978-3-642-18914-2_6} {\enquote {\bibinfo {title} {{Concepts in High
  Temperature Superconductivity}},}\ }\bibinfo {howpublished} {Springer Berlin
  Heidelberg},\ \bibinfo {address} {Berlin, Heidelberg} (\bibinfo {year}
  {2004})\BibitemShut {NoStop}%
\bibitem [{\citenamefont {Lee}\ \emph {et~al.}(2006)\citenamefont {Lee},
  \citenamefont {Nagaosa},\ and\ \citenamefont {Wen}}]{Lee06}%
  \BibitemOpen
  \bibfield  {author} {\bibinfo {author} {\bibfnamefont {P.~A.}\ \bibnamefont
  {Lee}}, \bibinfo {author} {\bibfnamefont {N.}~\bibnamefont {Nagaosa}}, \ and\
  \bibinfo {author} {\bibfnamefont {X.-G.}\ \bibnamefont {Wen}},\ }\href
  {\doibase 10.1103/RevModPhys.78.17} {\bibfield  {journal} {\bibinfo
  {journal} {Rev. Mod. Phys.}\ }\textbf {\bibinfo {volume} {78}},\ \bibinfo
  {pages} {17} (\bibinfo {year} {2006})}\BibitemShut {NoStop}%
\bibitem [{\citenamefont {Le~Hur}\ and\ \citenamefont
  {Rice}(2009)}]{LeHur:2009iw}%
  \BibitemOpen
  \bibfield  {author} {\bibinfo {author} {\bibfnamefont {K.}~\bibnamefont
  {Le~Hur}}\ and\ \bibinfo {author} {\bibfnamefont {T.~M.}\ \bibnamefont
  {Rice}},\ }\href {\doibase 10.1016/j.aop.2009.02.004} {\bibfield  {journal}
  {\bibinfo  {journal} {Annals of Physics}\ }\textbf {\bibinfo {volume}
  {324}},\ \bibinfo {pages} {1452} (\bibinfo {year} {2009})}\BibitemShut
  {NoStop}%
\bibitem [{\citenamefont {Rice}\ \emph {et~al.}(2012)\citenamefont {Rice},
  \citenamefont {Yang},\ and\ \citenamefont {Zhang}}]{Rice12}%
  \BibitemOpen
  \bibfield  {author} {\bibinfo {author} {\bibfnamefont {T.~M.}\ \bibnamefont
  {Rice}}, \bibinfo {author} {\bibfnamefont {K.-Y.}\ \bibnamefont {Yang}}, \
  and\ \bibinfo {author} {\bibfnamefont {F.~C.}\ \bibnamefont {Zhang}},\ }\href
  {\doibase 10.1088/0034-4885/75/1/016502} {\bibfield  {journal} {\bibinfo
  {journal} {Rep. Prog. Phys.}\ }\textbf {\bibinfo {volume} {75}},\ \bibinfo
  {pages} {016502} (\bibinfo {year} {2012})}\BibitemShut {NoStop}%
\bibitem [{\citenamefont {Norman}\ and\ \citenamefont
  {Proust}(2014)}]{Norman14}%
  \BibitemOpen
  \bibfield  {author} {\bibinfo {author} {\bibfnamefont {M.~R.}\ \bibnamefont
  {Norman}}\ and\ \bibinfo {author} {\bibfnamefont {C.}~\bibnamefont
  {Proust}},\ }\href {http://stacks.iop.org/1367-2630/16/i=4/a=045004}
  {\bibfield  {journal} {\bibinfo  {journal} {New J. Phys.}\ }\textbf {\bibinfo
  {volume} {16}},\ \bibinfo {pages} {045004} (\bibinfo {year}
  {2014})}\BibitemShut {NoStop}%
\bibitem [{\citenamefont {Carbotte}\ \emph {et~al.}(2011)\citenamefont
  {Carbotte}, \citenamefont {Timusk},\ and\ \citenamefont
  {Hwang}}]{Carbotte:2011ip}%
  \BibitemOpen
  \bibfield  {author} {\bibinfo {author} {\bibfnamefont {J.~P.}\ \bibnamefont
  {Carbotte}}, \bibinfo {author} {\bibfnamefont {T.}~\bibnamefont {Timusk}}, \
  and\ \bibinfo {author} {\bibfnamefont {J.}~\bibnamefont {Hwang}},\ }\href
  {\doibase 10.1088/0034-4885/74/6/066501} {\bibfield  {journal} {\bibinfo
  {journal} {Rep. Prog. Phys.}\ }\textbf {\bibinfo {volume} {74}},\ \bibinfo
  {pages} {066501} (\bibinfo {year} {2011})}\BibitemShut {NoStop}%
\bibitem [{\citenamefont {Eschrig}(2006)}]{Eschrig:2006ky}%
  \BibitemOpen
  \bibfield  {author} {\bibinfo {author} {\bibfnamefont {M.}~\bibnamefont
  {Eschrig}},\ }\href {\doibase 10.1080/00018730600645636} {\bibfield
  {journal} {\bibinfo  {journal} {Advances in Physics}\ }\textbf {\bibinfo
  {volume} {55}},\ \bibinfo {pages} {47} (\bibinfo {year} {2006})}\BibitemShut
  {NoStop}%
\bibitem [{\citenamefont {Fradkin}\ \emph
  {et~al.}(2015{\natexlab{a}})\citenamefont {Fradkin}, \citenamefont
  {Kivelson},\ and\ \citenamefont {Tranquada}}]{Fradkin:2015ch}%
  \BibitemOpen
  \bibfield  {author} {\bibinfo {author} {\bibfnamefont {E.}~\bibnamefont
  {Fradkin}}, \bibinfo {author} {\bibfnamefont {S.~A.}\ \bibnamefont
  {Kivelson}}, \ and\ \bibinfo {author} {\bibfnamefont {J.~M.}\ \bibnamefont
  {Tranquada}},\ }\href {\doibase 10.1103/revmodphys.87.457} {\bibfield
  {journal} {\bibinfo  {journal} {Rev. Mod. Phys.}\ }\textbf {\bibinfo {volume}
  {87}},\ \bibinfo {pages} {457} (\bibinfo {year}
  {2015}{\natexlab{a}})}\BibitemShut {NoStop}%
\bibitem [{\citenamefont {Kloss}\ \emph {et~al.}(2016)\citenamefont {Kloss},
  \citenamefont {Montiel}, \citenamefont {de~Carvalho}, \citenamefont
  {Freire},\ and\ \citenamefont {P{\'e}pin}}]{Kloss:2016hu}%
  \BibitemOpen
  \bibfield  {author} {\bibinfo {author} {\bibfnamefont {T.}~\bibnamefont
  {Kloss}}, \bibinfo {author} {\bibfnamefont {X.}~\bibnamefont {Montiel}},
  \bibinfo {author} {\bibfnamefont {V.~S.}\ \bibnamefont {de~Carvalho}},
  \bibinfo {author} {\bibfnamefont {H.}~\bibnamefont {Freire}}, \ and\ \bibinfo
  {author} {\bibfnamefont {C.}~\bibnamefont {P{\'e}pin}},\ }\href
  {http://stacks.iop.org/0034-4885/79/i=8/a=084507?key=crossref.f96d8d872a758c55b6cbb75f83e9cea6}
  {\bibfield  {journal} {\bibinfo  {journal} {Rep. Prog. Phys.}\ }\textbf
  {\bibinfo {volume} {79}} (\bibinfo {year} {2016})}\BibitemShut {NoStop}%
\bibitem [{\citenamefont {Abanov}\ \emph {et~al.}(2003)\citenamefont {Abanov},
  \citenamefont {Chubukov},\ and\ \citenamefont {Schmalian}}]{abanov03}%
  \BibitemOpen
  \bibfield  {author} {\bibinfo {author} {\bibfnamefont {A.}~\bibnamefont
  {Abanov}}, \bibinfo {author} {\bibfnamefont {A.~V.}\ \bibnamefont
  {Chubukov}}, \ and\ \bibinfo {author} {\bibfnamefont {J.}~\bibnamefont
  {Schmalian}},\ }\href {\doibase 10.1080/0001873021000057123} {\bibfield
  {journal} {\bibinfo  {journal} {Adv. Phys.}\ }\textbf {\bibinfo {volume}
  {52}},\ \bibinfo {pages} {119} (\bibinfo {year} {2003})}\BibitemShut
  {NoStop}%
\bibitem [{\citenamefont {Chubukov}\ \emph {et~al.}(2008)\citenamefont
  {Chubukov}, \citenamefont {Pines},\ and\ \citenamefont {Schmalian}}]{sfbook}%
  \BibitemOpen
  \bibfield  {author} {\bibinfo {author} {\bibfnamefont {A.}~\bibnamefont
  {Chubukov}}, \bibinfo {author} {\bibfnamefont {D.}~\bibnamefont {Pines}}, \
  and\ \bibinfo {author} {\bibfnamefont {J.}~\bibnamefont {Schmalian}},\ }in\
  \href {\doibase 10.1007/978-3-540-73253-2_22} {\emph {\bibinfo {booktitle}
  {Superconductivity}}},\ \bibinfo {editor} {edited by\ \bibinfo {editor}
  {\bibfnamefont {K.}~\bibnamefont {Bennemann}}\ and\ \bibinfo {editor}
  {\bibfnamefont {J.}~\bibnamefont {Ketterson}}}\ (\bibinfo  {publisher}
  {Springer Berlin Heidelberg},\ \bibinfo {year} {2008})\BibitemShut {NoStop}%
\bibitem [{\citenamefont {Sorella}\ \emph {et~al.}(2002)\citenamefont
  {Sorella}, \citenamefont {Martins}, \citenamefont {Becca}, \citenamefont
  {Gazza}, \citenamefont {Capriotti}, \citenamefont {Parola},\ and\
  \citenamefont {Dagotto}}]{Sorella02}%
  \BibitemOpen
  \bibfield  {author} {\bibinfo {author} {\bibfnamefont {S.}~\bibnamefont
  {Sorella}}, \bibinfo {author} {\bibfnamefont {G.~B.}\ \bibnamefont
  {Martins}}, \bibinfo {author} {\bibfnamefont {F.}~\bibnamefont {Becca}},
  \bibinfo {author} {\bibfnamefont {C.}~\bibnamefont {Gazza}}, \bibinfo
  {author} {\bibfnamefont {L.}~\bibnamefont {Capriotti}}, \bibinfo {author}
  {\bibfnamefont {A.}~\bibnamefont {Parola}}, \ and\ \bibinfo {author}
  {\bibfnamefont {E.}~\bibnamefont {Dagotto}},\ }\href {\doibase
  10.1103/PhysRevLett.88.117002} {\bibfield  {journal} {\bibinfo  {journal}
  {Phys. Rev. Lett.}\ }\textbf {\bibinfo {volume} {88}},\ \bibinfo {pages}
  {117002} (\bibinfo {year} {2002})}\BibitemShut {NoStop}%
\bibitem [{\citenamefont {Gull}\ \emph {et~al.}(2013)\citenamefont {Gull},
  \citenamefont {Parcollet},\ and\ \citenamefont {Millis}}]{Gull:2013hh}%
  \BibitemOpen
  \bibfield  {author} {\bibinfo {author} {\bibfnamefont {E.}~\bibnamefont
  {Gull}}, \bibinfo {author} {\bibfnamefont {O.}~\bibnamefont {Parcollet}}, \
  and\ \bibinfo {author} {\bibfnamefont {A.~J.}\ \bibnamefont {Millis}},\
  }\href {\doibase 10.1103/PhysRevLett.110.216405} {\bibfield  {journal}
  {\bibinfo  {journal} {Phys. Rev. Lett.}\ }\textbf {\bibinfo {volume} {110}},\
  \bibinfo {pages} {216405} (\bibinfo {year} {2013})}\BibitemShut {NoStop}%
\bibitem [{\citenamefont {Wang}\ \emph {et~al.}(1990)\citenamefont {Wang},
  \citenamefont {Kotliar},\ and\ \citenamefont {Wang}}]{Kotliar90}%
  \BibitemOpen
  \bibfield  {author} {\bibinfo {author} {\bibfnamefont {Z.}~\bibnamefont
  {Wang}}, \bibinfo {author} {\bibfnamefont {G.}~\bibnamefont {Kotliar}}, \
  and\ \bibinfo {author} {\bibfnamefont {X.-F.}\ \bibnamefont {Wang}},\ }\href
  {\doibase 10.1103/PhysRevB.42.8690} {\bibfield  {journal} {\bibinfo
  {journal} {Phys. Rev. B}\ }\textbf {\bibinfo {volume} {42}},\ \bibinfo
  {pages} {8690} (\bibinfo {year} {1990})}\BibitemShut {NoStop}%
\bibitem [{\citenamefont {Varma}(1997)}]{Varma97}%
  \BibitemOpen
  \bibfield  {author} {\bibinfo {author} {\bibfnamefont {C.~M.}\ \bibnamefont
  {Varma}},\ }\href {\doibase 10.1103/PhysRevB.55.14554} {\bibfield  {journal}
  {\bibinfo  {journal} {Phys. Rev. B}\ }\textbf {\bibinfo {volume} {55}},\
  \bibinfo {pages} {14554} (\bibinfo {year} {1997})}\BibitemShut {NoStop}%
\bibitem [{\citenamefont {Zhang}(1997)}]{Zhang97}%
  \BibitemOpen
  \bibfield  {author} {\bibinfo {author} {\bibfnamefont {S.-C.}\ \bibnamefont
  {Zhang}},\ }\href {\doibase 10.1126/science.275.5303.1089} {\bibfield
  {journal} {\bibinfo  {journal} {Science}\ }\textbf {\bibinfo {volume}
  {275}},\ \bibinfo {pages} {1089} (\bibinfo {year} {1997})}\BibitemShut
  {NoStop}%
\bibitem [{\citenamefont {Demler}\ \emph {et~al.}(2004)\citenamefont {Demler},
  \citenamefont {Hanke},\ and\ \citenamefont {Zhang}}]{Demler04}%
  \BibitemOpen
  \bibfield  {author} {\bibinfo {author} {\bibfnamefont {E.}~\bibnamefont
  {Demler}}, \bibinfo {author} {\bibfnamefont {W.}~\bibnamefont {Hanke}}, \
  and\ \bibinfo {author} {\bibfnamefont {S.-C.}\ \bibnamefont {Zhang}},\ }\href
  {\doibase 10.1103/RevModPhys.76.909} {\bibfield  {journal} {\bibinfo
  {journal} {Rev. Mod. Phys.}\ }\textbf {\bibinfo {volume} {76}},\ \bibinfo
  {pages} {909} (\bibinfo {year} {2004})}\BibitemShut {NoStop}%
\bibitem [{\citenamefont {{Montiel}}\ \emph {et~al.}(2015)\citenamefont
  {{Montiel}}, \citenamefont {{Kloss}},\ and\ \citenamefont
  {{P{\'e}pin}}}]{Kloss15a}%
  \BibitemOpen
  \bibfield  {author} {\bibinfo {author} {\bibfnamefont {X.}~\bibnamefont
  {{Montiel}}}, \bibinfo {author} {\bibfnamefont {T.}~\bibnamefont {{Kloss}}},
  \ and\ \bibinfo {author} {\bibfnamefont {C.}~\bibnamefont {{P{\'e}pin}}},\
  }\href {http://arxiv.org/abs/1510.03038} {\bibfield  {journal} {\bibinfo
  {journal} {arXiv:1510.03038}\ } (\bibinfo {year} {2015})}\BibitemShut
  {NoStop}%
\bibitem [{\citenamefont {Montiel}\ \emph
  {et~al.}(2017{\natexlab{a}})\citenamefont {Montiel}, \citenamefont {Kloss},\
  and\ \citenamefont {P\'epin}}]{Montiel2017}%
  \BibitemOpen
  \bibfield  {author} {\bibinfo {author} {\bibfnamefont {X.}~\bibnamefont
  {Montiel}}, \bibinfo {author} {\bibfnamefont {T.}~\bibnamefont {Kloss}}, \
  and\ \bibinfo {author} {\bibfnamefont {C.}~\bibnamefont {P\'epin}},\ }\href
  {\doibase 10.1103/PhysRevB.95.104510} {\bibfield  {journal} {\bibinfo
  {journal} {Phys. Rev. B}\ }\textbf {\bibinfo {volume} {95}},\ \bibinfo
  {pages} {104510} (\bibinfo {year} {2017}{\natexlab{a}})}\BibitemShut
  {NoStop}%
\bibitem [{\citenamefont {Badoux}\ \emph {et~al.}(2015)\citenamefont {Badoux},
  \citenamefont {Tabis}, \citenamefont {Lalibert{\'{e}}}, \citenamefont
  {Grissonnanche}, \citenamefont {Vignolle}, \citenamefont {Vignolles},
  \citenamefont {B{\'{e}}ard}, \citenamefont {Bonn}, \citenamefont {Hardy},
  \citenamefont {Liang}, \citenamefont {Doiron-Leyraud}, \citenamefont
  {Taillefer},\ and\ \citenamefont {Proust}}]{Badoux2015}%
  \BibitemOpen
  \bibfield  {author} {\bibinfo {author} {\bibfnamefont {S.}~\bibnamefont
  {Badoux}}, \bibinfo {author} {\bibfnamefont {W.}~\bibnamefont {Tabis}},
  \bibinfo {author} {\bibfnamefont {F.}~\bibnamefont {Lalibert{\'{e}}}},
  \bibinfo {author} {\bibfnamefont {G.}~\bibnamefont {Grissonnanche}}, \bibinfo
  {author} {\bibfnamefont {B.}~\bibnamefont {Vignolle}}, \bibinfo {author}
  {\bibfnamefont {D.}~\bibnamefont {Vignolles}}, \bibinfo {author}
  {\bibfnamefont {J.}~\bibnamefont {B{\'{e}}ard}}, \bibinfo {author}
  {\bibfnamefont {D.~A.}\ \bibnamefont {Bonn}}, \bibinfo {author}
  {\bibfnamefont {W.~N.}\ \bibnamefont {Hardy}}, \bibinfo {author}
  {\bibfnamefont {R.}~\bibnamefont {Liang}}, \bibinfo {author} {\bibfnamefont
  {N.}~\bibnamefont {Doiron-Leyraud}}, \bibinfo {author} {\bibfnamefont
  {L.}~\bibnamefont {Taillefer}}, \ and\ \bibinfo {author} {\bibfnamefont
  {C.}~\bibnamefont {Proust}},\ }\href {\doibase 10.1038/nature16983}
  {\bibfield  {journal} {\bibinfo  {journal} {Nature}\ }\textbf {\bibinfo
  {volume} {531}},\ \bibinfo {pages} {210} (\bibinfo {year}
  {2015})}\BibitemShut {NoStop}%
\bibitem [{\citenamefont {Collignon}\ \emph {et~al.}()\citenamefont
  {Collignon}, \citenamefont {Badoux}, \citenamefont {Afshar}, \citenamefont
  {Michon}, \citenamefont {Laliberte}, \citenamefont {Cyr-Choiniere},
  \citenamefont {Zhou}, \citenamefont {Licciardello}, \citenamefont {Wiedmann},
  \citenamefont {Doiron-Leyraud},\ and\ \citenamefont
  {Taillefer}}]{Collignon2016}%
  \BibitemOpen
  \bibfield  {author} {\bibinfo {author} {\bibfnamefont {C.}~\bibnamefont
  {Collignon}}, \bibinfo {author} {\bibfnamefont {S.}~\bibnamefont {Badoux}},
  \bibinfo {author} {\bibfnamefont {S.~A.~A.}\ \bibnamefont {Afshar}}, \bibinfo
  {author} {\bibfnamefont {B.}~\bibnamefont {Michon}}, \bibinfo {author}
  {\bibfnamefont {F.}~\bibnamefont {Laliberte}}, \bibinfo {author}
  {\bibfnamefont {O.}~\bibnamefont {Cyr-Choiniere}}, \bibinfo {author}
  {\bibfnamefont {J.~S.}\ \bibnamefont {Zhou}}, \bibinfo {author}
  {\bibfnamefont {S.}~\bibnamefont {Licciardello}}, \bibinfo {author}
  {\bibfnamefont {S.}~\bibnamefont {Wiedmann}}, \bibinfo {author}
  {\bibfnamefont {N.}~\bibnamefont {Doiron-Leyraud}}, \ and\ \bibinfo {author}
  {\bibfnamefont {L.}~\bibnamefont {Taillefer}},\ }\href
  {http://arxiv.org/abs/1607.05693} {\ }\Eprint
  {http://arxiv.org/abs/1607.05693} {arXiv:1607.05693} \BibitemShut {NoStop}%
\bibitem [{\citenamefont {{Laliberte}}\ \emph {et~al.}(2016)\citenamefont
  {{Laliberte}}, \citenamefont {{Tabis}}, \citenamefont {{Badoux}},
  \citenamefont {{Vignolle}}, \citenamefont {{Destraz}}, \citenamefont
  {{Momono}}, \citenamefont {{Kurosawa}}, \citenamefont {{Yamada}},
  \citenamefont {{Takagi}}, \citenamefont {{Doiron-Leyraud}}, \citenamefont
  {{Proust}},\ and\ \citenamefont {{Taillefer}}}]{Laliberte2016}%
  \BibitemOpen
  \bibfield  {author} {\bibinfo {author} {\bibfnamefont {F.}~\bibnamefont
  {{Laliberte}}}, \bibinfo {author} {\bibfnamefont {W.}~\bibnamefont
  {{Tabis}}}, \bibinfo {author} {\bibfnamefont {S.}~\bibnamefont {{Badoux}}},
  \bibinfo {author} {\bibfnamefont {B.}~\bibnamefont {{Vignolle}}}, \bibinfo
  {author} {\bibfnamefont {D.}~\bibnamefont {{Destraz}}}, \bibinfo {author}
  {\bibfnamefont {N.}~\bibnamefont {{Momono}}}, \bibinfo {author}
  {\bibfnamefont {T.}~\bibnamefont {{Kurosawa}}}, \bibinfo {author}
  {\bibfnamefont {K.}~\bibnamefont {{Yamada}}}, \bibinfo {author}
  {\bibfnamefont {H.}~\bibnamefont {{Takagi}}}, \bibinfo {author}
  {\bibfnamefont {N.}~\bibnamefont {{Doiron-Leyraud}}}, \bibinfo {author}
  {\bibfnamefont {C.}~\bibnamefont {{Proust}}}, \ and\ \bibinfo {author}
  {\bibfnamefont {L.}~\bibnamefont {{Taillefer}}},\ }\href@noop {} {\bibfield
  {journal} {\bibinfo  {journal} {ArXiv e-prints}\ } (\bibinfo {year}
  {2016})},\ \Eprint {http://arxiv.org/abs/1606.04491} {arXiv:1606.04491}
  \BibitemShut {NoStop}%
\bibitem [{\citenamefont {Badoux}\ \emph {et~al.}(2016)\citenamefont {Badoux},
  \citenamefont {Afshar}, \citenamefont {Michon}, \citenamefont {Ouellet},
  \citenamefont {Fortier}, \citenamefont {LeBoeuf}, \citenamefont {Croft},
  \citenamefont {Lester}, \citenamefont {Hayden}, \citenamefont {Takagi},
  \citenamefont {Yamada}, \citenamefont {Graf}, \citenamefont
  {Doiron-Leyraud},\ and\ \citenamefont {Taillefer}}]{Badoux:2016kg}%
  \BibitemOpen
  \bibfield  {author} {\bibinfo {author} {\bibfnamefont {S.}~\bibnamefont
  {Badoux}}, \bibinfo {author} {\bibfnamefont {S.~A.~A.}\ \bibnamefont
  {Afshar}}, \bibinfo {author} {\bibfnamefont {B.}~\bibnamefont {Michon}},
  \bibinfo {author} {\bibfnamefont {A.}~\bibnamefont {Ouellet}}, \bibinfo
  {author} {\bibfnamefont {S.}~\bibnamefont {Fortier}}, \bibinfo {author}
  {\bibfnamefont {D.}~\bibnamefont {LeBoeuf}}, \bibinfo {author} {\bibfnamefont
  {T.~P.}\ \bibnamefont {Croft}}, \bibinfo {author} {\bibfnamefont
  {C.}~\bibnamefont {Lester}}, \bibinfo {author} {\bibfnamefont {S.~M.}\
  \bibnamefont {Hayden}}, \bibinfo {author} {\bibfnamefont {H.}~\bibnamefont
  {Takagi}}, \bibinfo {author} {\bibfnamefont {K.}~\bibnamefont {Yamada}},
  \bibinfo {author} {\bibfnamefont {D.}~\bibnamefont {Graf}}, \bibinfo {author}
  {\bibfnamefont {N.}~\bibnamefont {Doiron-Leyraud}}, \ and\ \bibinfo {author}
  {\bibfnamefont {L.}~\bibnamefont {Taillefer}},\ }\href
  {http://link.aps.org/doi/10.1103/PhysRevX.6.021004} {\bibfield  {journal}
  {\bibinfo  {journal} {Physical Review X}\ }\textbf {\bibinfo {volume} {6}},\
  \bibinfo {pages} {021004} (\bibinfo {year} {2016})}\BibitemShut {NoStop}%
\bibitem [{\citenamefont {Shen}\ \emph {et~al.}(2005)\citenamefont {Shen},
  \citenamefont {Ronning}, \citenamefont {Lu}, \citenamefont {Baumberger},
  \citenamefont {Ingle}, \citenamefont {Lee}, \citenamefont {Meevasana},
  \citenamefont {Kohsaka}, \citenamefont {Azuma}, \citenamefont {Takano},
  \citenamefont {Takagi},\ and\ \citenamefont {Shen}}]{Shen:2005ir}%
  \BibitemOpen
  \bibfield  {author} {\bibinfo {author} {\bibfnamefont {K.~M.}\ \bibnamefont
  {Shen}}, \bibinfo {author} {\bibfnamefont {F.}~\bibnamefont {Ronning}},
  \bibinfo {author} {\bibfnamefont {D.~H.}\ \bibnamefont {Lu}}, \bibinfo
  {author} {\bibfnamefont {F.}~\bibnamefont {Baumberger}}, \bibinfo {author}
  {\bibfnamefont {N.~J.~C.}\ \bibnamefont {Ingle}}, \bibinfo {author}
  {\bibfnamefont {W.~S.}\ \bibnamefont {Lee}}, \bibinfo {author} {\bibfnamefont
  {W.}~\bibnamefont {Meevasana}}, \bibinfo {author} {\bibfnamefont
  {Y.}~\bibnamefont {Kohsaka}}, \bibinfo {author} {\bibfnamefont
  {M.}~\bibnamefont {Azuma}}, \bibinfo {author} {\bibfnamefont
  {M.}~\bibnamefont {Takano}}, \bibinfo {author} {\bibfnamefont
  {H.}~\bibnamefont {Takagi}}, \ and\ \bibinfo {author} {\bibfnamefont {Z.-X.}\
  \bibnamefont {Shen}},\ }\href {\doibase 10.1126/science.1103627} {\bibfield
  {journal} {\bibinfo  {journal} {Science}\ }\textbf {\bibinfo {volume}
  {307}},\ \bibinfo {pages} {901} (\bibinfo {year} {2005})}\BibitemShut
  {NoStop}%
\bibitem [{\citenamefont {Vignolle}\ \emph {et~al.}(2008)\citenamefont
  {Vignolle}, \citenamefont {Carrington}, \citenamefont {Cooper}, \citenamefont
  {French}, \citenamefont {Mackenzie}, \citenamefont {Jaudet}, \citenamefont
  {Vignolles}, \citenamefont {Proust},\ and\ \citenamefont
  {Hussey}}]{Vignolle2008}%
  \BibitemOpen
  \bibfield  {author} {\bibinfo {author} {\bibfnamefont {B.}~\bibnamefont
  {Vignolle}}, \bibinfo {author} {\bibfnamefont {a.}~\bibnamefont
  {Carrington}}, \bibinfo {author} {\bibfnamefont {R.~a.}\ \bibnamefont
  {Cooper}}, \bibinfo {author} {\bibfnamefont {M.~M.~J.}\ \bibnamefont
  {French}}, \bibinfo {author} {\bibfnamefont {a.~P.}\ \bibnamefont
  {Mackenzie}}, \bibinfo {author} {\bibfnamefont {C.}~\bibnamefont {Jaudet}},
  \bibinfo {author} {\bibfnamefont {D.}~\bibnamefont {Vignolles}}, \bibinfo
  {author} {\bibfnamefont {C.}~\bibnamefont {Proust}}, \ and\ \bibinfo {author}
  {\bibfnamefont {N.~E.}\ \bibnamefont {Hussey}},\ }\href {\doibase
  10.1038/nature07323} {\bibfield  {journal} {\bibinfo  {journal} {Nature}\
  }\textbf {\bibinfo {volume} {455}},\ \bibinfo {pages} {952} (\bibinfo {year}
  {2008})}\BibitemShut {NoStop}%
\bibitem [{\citenamefont {Sebastian}\ \emph {et~al.}(2011)\citenamefont
  {Sebastian}, \citenamefont {Harrison},\ and\ \citenamefont
  {Lonzarich}}]{Sebastian2011}%
  \BibitemOpen
  \bibfield  {author} {\bibinfo {author} {\bibfnamefont {S.~E.}\ \bibnamefont
  {Sebastian}}, \bibinfo {author} {\bibfnamefont {N.}~\bibnamefont {Harrison}},
  \ and\ \bibinfo {author} {\bibfnamefont {G.~G.}\ \bibnamefont {Lonzarich}},\
  }\href {\doibase 10.1098/rsta.2010.0243} {\bibfield  {journal} {\bibinfo
  {journal} {Phil. Trans. R. Soc. A}\ }\textbf {\bibinfo {volume} {369}},\
  \bibinfo {pages} {1687} (\bibinfo {year} {2011})}\BibitemShut {NoStop}%
\bibitem [{\citenamefont {Balakirev}\ \emph {et~al.}(2003)\citenamefont
  {Balakirev}, \citenamefont {Betts}, \citenamefont {Migliori}, \citenamefont
  {Ono}, \citenamefont {Ando},\ and\ \citenamefont
  {Boebinger}}]{Balakirev2003}%
  \BibitemOpen
  \bibfield  {author} {\bibinfo {author} {\bibfnamefont {F.~F.}\ \bibnamefont
  {Balakirev}}, \bibinfo {author} {\bibfnamefont {J.~B.}\ \bibnamefont
  {Betts}}, \bibinfo {author} {\bibfnamefont {A.}~\bibnamefont {Migliori}},
  \bibinfo {author} {\bibfnamefont {S.}~\bibnamefont {Ono}}, \bibinfo {author}
  {\bibfnamefont {Y.}~\bibnamefont {Ando}}, \ and\ \bibinfo {author}
  {\bibfnamefont {G.~S.}\ \bibnamefont {Boebinger}},\ }\href {\doibase
  10.1038/nature01890} {\bibfield  {journal} {\bibinfo  {journal} {Nature}\
  }\textbf {\bibinfo {volume} {424}},\ \bibinfo {pages} {912} (\bibinfo {year}
  {2003})}\BibitemShut {NoStop}%
\bibitem [{\citenamefont {Balakirev}\ \emph {et~al.}(2009)\citenamefont
  {Balakirev}, \citenamefont {Betts}, \citenamefont {Migliori}, \citenamefont
  {Tsukada}, \citenamefont {Ando},\ and\ \citenamefont
  {Boebinger}}]{Balakirev2009}%
  \BibitemOpen
  \bibfield  {author} {\bibinfo {author} {\bibfnamefont {F.~F.}\ \bibnamefont
  {Balakirev}}, \bibinfo {author} {\bibfnamefont {J.~B.}\ \bibnamefont
  {Betts}}, \bibinfo {author} {\bibfnamefont {A.}~\bibnamefont {Migliori}},
  \bibinfo {author} {\bibfnamefont {I.}~\bibnamefont {Tsukada}}, \bibinfo
  {author} {\bibfnamefont {Y.}~\bibnamefont {Ando}}, \ and\ \bibinfo {author}
  {\bibfnamefont {G.~S.}\ \bibnamefont {Boebinger}},\ }\href {\doibase
  10.1103/PhysRevLett.102.017004} {\bibfield  {journal} {\bibinfo  {journal}
  {Phys. Rev. Lett.}\ }\textbf {\bibinfo {volume} {102}},\ \bibinfo {pages}
  {017004} (\bibinfo {year} {2009})}\BibitemShut {NoStop}%
\bibitem [{\citenamefont {Sachdev}\ \emph {et~al.}(2016)\citenamefont
  {Sachdev}, \citenamefont {Berg}, \citenamefont {Chatterjee},\ and\
  \citenamefont {Schattner}}]{Sachdev2016}%
  \BibitemOpen
  \bibfield  {author} {\bibinfo {author} {\bibfnamefont {S.}~\bibnamefont
  {Sachdev}}, \bibinfo {author} {\bibfnamefont {E.}~\bibnamefont {Berg}},
  \bibinfo {author} {\bibfnamefont {S.}~\bibnamefont {Chatterjee}}, \ and\
  \bibinfo {author} {\bibfnamefont {Y.}~\bibnamefont {Schattner}},\ }\href
  {\doibase 10.1103/PhysRevB.94.115147} {\bibfield  {journal} {\bibinfo
  {journal} {Phys. Rev. B}\ }\textbf {\bibinfo {volume} {94}},\ \bibinfo
  {pages} {115147} (\bibinfo {year} {2016})}\BibitemShut {NoStop}%
\bibitem [{\citenamefont {Zou}\ \emph {et~al.}(2017)\citenamefont {Zou},
  \citenamefont {Lederer},\ and\ \citenamefont {Senthil}}]{Zou2017}%
  \BibitemOpen
  \bibfield  {author} {\bibinfo {author} {\bibfnamefont {L.}~\bibnamefont
  {Zou}}, \bibinfo {author} {\bibfnamefont {S.}~\bibnamefont {Lederer}}, \ and\
  \bibinfo {author} {\bibfnamefont {T.}~\bibnamefont {Senthil}},\ }\href
  {\doibase 10.1103/PhysRevB.95.245135} {\bibfield  {journal} {\bibinfo
  {journal} {Phys. Rev. B}\ }\textbf {\bibinfo {volume} {95}},\ \bibinfo
  {pages} {245135} (\bibinfo {year} {2017})}\BibitemShut {NoStop}%
\bibitem [{\citenamefont {{Sharma}}\ \emph {et~al.}(2017)\citenamefont
  {{Sharma}}, \citenamefont {{Nandy}}, \citenamefont {{Taraphder}},\ and\
  \citenamefont {{Tewari}}}]{Sharma2017}%
  \BibitemOpen
  \bibfield  {author} {\bibinfo {author} {\bibfnamefont {G.}~\bibnamefont
  {{Sharma}}}, \bibinfo {author} {\bibfnamefont {S.}~\bibnamefont {{Nandy}}},
  \bibinfo {author} {\bibfnamefont {A.}~\bibnamefont {{Taraphder}}}, \ and\
  \bibinfo {author} {\bibfnamefont {S.}~\bibnamefont {{Tewari}}},\ }\href@noop
  {} {\bibfield  {journal} {\bibinfo  {journal} {ArXiv e-prints}\ } (\bibinfo
  {year} {2017})},\ \Eprint {http://arxiv.org/abs/1703.04620}
  {arXiv:1703.04620} \BibitemShut {NoStop}%
\bibitem [{\citenamefont {Emery}\ and\ \citenamefont
  {Kivelson}(1995)}]{Emery1995}%
  \BibitemOpen
  \bibfield  {author} {\bibinfo {author} {\bibfnamefont {V.~J.}\ \bibnamefont
  {Emery}}\ and\ \bibinfo {author} {\bibfnamefont {S.~a.}\ \bibnamefont
  {Kivelson}},\ }\href {\doibase 10.1038/374434a0} {\bibfield  {journal}
  {\bibinfo  {journal} {Nature}\ }\textbf {\bibinfo {volume} {374}},\ \bibinfo
  {pages} {434} (\bibinfo {year} {1995})}\BibitemShut {NoStop}%
\bibitem [{\citenamefont {Kanigel}\ \emph {et~al.}(2008)\citenamefont
  {Kanigel}, \citenamefont {Chatterjee}, \citenamefont {Randeria},
  \citenamefont {Norman}, \citenamefont {Koren}, \citenamefont {Kadowaki},\
  and\ \citenamefont {Campuzano}}]{Kanigel:2008wm}%
  \BibitemOpen
  \bibfield  {author} {\bibinfo {author} {\bibfnamefont {A.}~\bibnamefont
  {Kanigel}}, \bibinfo {author} {\bibfnamefont {U.}~\bibnamefont {Chatterjee}},
  \bibinfo {author} {\bibfnamefont {M.}~\bibnamefont {Randeria}}, \bibinfo
  {author} {\bibfnamefont {M.~R.}\ \bibnamefont {Norman}}, \bibinfo {author}
  {\bibfnamefont {G.}~\bibnamefont {Koren}}, \bibinfo {author} {\bibfnamefont
  {K.}~\bibnamefont {Kadowaki}}, \ and\ \bibinfo {author} {\bibfnamefont
  {J.~C.}\ \bibnamefont {Campuzano}},\ }\href
  {http://prl.aps.org/abstract/PRL/v101/i13/e137002} {\bibfield  {journal}
  {\bibinfo  {journal} {Phys. Rev. Lett.}\ }\textbf {\bibinfo {volume} {101}},\
  \bibinfo {pages} {137002} (\bibinfo {year} {2008})}\BibitemShut {NoStop}%
\bibitem [{\citenamefont {Yang}\ \emph
  {et~al.}(2006{\natexlab{a}})\citenamefont {Yang}, \citenamefont {Rice},\ and\
  \citenamefont {Zhang}}]{Yang2006}%
  \BibitemOpen
  \bibfield  {author} {\bibinfo {author} {\bibfnamefont {K.~Y.}\ \bibnamefont
  {Yang}}, \bibinfo {author} {\bibfnamefont {T.~M.}\ \bibnamefont {Rice}}, \
  and\ \bibinfo {author} {\bibfnamefont {F.~C.}\ \bibnamefont {Zhang}},\ }\href
  {\doibase 10.1103/PhysRevB.73.174501} {\bibfield  {journal} {\bibinfo
  {journal} {Phys. Rev. B}\ }\textbf {\bibinfo {volume} {73}},\ \bibinfo
  {pages} {174501} (\bibinfo {year} {2006}{\natexlab{a}})}\BibitemShut
  {NoStop}%
\bibitem [{\citenamefont {Storey}(2016)}]{Storey2016}%
  \BibitemOpen
  \bibfield  {author} {\bibinfo {author} {\bibfnamefont {J.~G.}\ \bibnamefont
  {Storey}},\ }\href {\doibase 10.1209/0295-5075/113/27003} {\bibfield
  {journal} {\bibinfo  {journal} {EPL (Europhysics Lett.)}\ }\textbf {\bibinfo
  {volume} {113}},\ \bibinfo {pages} {27003} (\bibinfo {year}
  {2016})}\BibitemShut {NoStop}%
\bibitem [{\citenamefont {{Chatterjee}}\ \emph {et~al.}()\citenamefont
  {{Chatterjee}}, \citenamefont {{Sachdev}},\ and\ \citenamefont
  {{Eberlein}}}]{Chatterjee2017}%
  \BibitemOpen
  \bibfield  {author} {\bibinfo {author} {\bibfnamefont {S.}~\bibnamefont
  {{Chatterjee}}}, \bibinfo {author} {\bibfnamefont {S.}~\bibnamefont
  {{Sachdev}}}, \ and\ \bibinfo {author} {\bibfnamefont {A.}~\bibnamefont
  {{Eberlein}}},\ }\href@noop {} {\ }\Eprint {http://arxiv.org/abs/1704.02329}
  {arXiv:1704.02329} \BibitemShut {NoStop}%
\bibitem [{\citenamefont {Caprara}\ \emph {et~al.}(2016)\citenamefont
  {Caprara}, \citenamefont {Grilli}, \citenamefont {{Di Castro}},\ and\
  \citenamefont {Seibold}}]{Caprara2016}%
  \BibitemOpen
  \bibfield  {author} {\bibinfo {author} {\bibfnamefont {S.}~\bibnamefont
  {Caprara}}, \bibinfo {author} {\bibfnamefont {M.}~\bibnamefont {Grilli}},
  \bibinfo {author} {\bibfnamefont {C.}~\bibnamefont {{Di Castro}}}, \ and\
  \bibinfo {author} {\bibfnamefont {G.}~\bibnamefont {Seibold}},\ }\href
  {\doibase 10.1007/s10948-016-3775-9} {\bibfield  {journal} {\bibinfo
  {journal} {J. Supercond. Nov. Magn.}\ }\textbf {\bibinfo {volume} {30}},\
  \bibinfo {pages} {25} (\bibinfo {year} {2016})}\BibitemShut {NoStop}%
\bibitem [{\citenamefont {Fradkin}\ \emph
  {et~al.}(2015{\natexlab{b}})\citenamefont {Fradkin}, \citenamefont
  {Kivelson},\ and\ \citenamefont {Tranquada}}]{Fradkin15}%
  \BibitemOpen
  \bibfield  {author} {\bibinfo {author} {\bibfnamefont {E.}~\bibnamefont
  {Fradkin}}, \bibinfo {author} {\bibfnamefont {S.~A.}\ \bibnamefont
  {Kivelson}}, \ and\ \bibinfo {author} {\bibfnamefont {J.~M.}\ \bibnamefont
  {Tranquada}},\ }\href {\doibase 10.1103/RevModPhys.87.457} {\bibfield
  {journal} {\bibinfo  {journal} {Rev. Mod. Phys.}\ }\textbf {\bibinfo {volume}
  {87}},\ \bibinfo {pages} {457} (\bibinfo {year}
  {2015}{\natexlab{b}})}\BibitemShut {NoStop}%
\bibitem [{\citenamefont {{Morice}}\ \emph {et~al.}(2017)\citenamefont
  {{Morice}}, \citenamefont {{Chakraborty}}, \citenamefont {{Montiel}},\ and\
  \citenamefont {{P{\'e}pin}}}]{Morice2017skyrmions}%
  \BibitemOpen
  \bibfield  {author} {\bibinfo {author} {\bibfnamefont {C.}~\bibnamefont
  {{Morice}}}, \bibinfo {author} {\bibfnamefont {D.}~\bibnamefont
  {{Chakraborty}}}, \bibinfo {author} {\bibfnamefont {X.}~\bibnamefont
  {{Montiel}}}, \ and\ \bibinfo {author} {\bibfnamefont {C.}~\bibnamefont
  {{P{\'e}pin}}},\ }\href@noop {} {\bibfield  {journal} {\bibinfo  {journal}
  {ArXiv e-prints}\ } (\bibinfo {year} {2017})},\ \Eprint
  {http://arxiv.org/abs/1707.08497} {arXiv:1707.08497} \BibitemShut {NoStop}%
\bibitem [{\citenamefont {Vishik}\ \emph {et~al.}(2012)\citenamefont {Vishik},
  \citenamefont {Hashimoto}, \citenamefont {He}, \citenamefont {Lee},
  \citenamefont {Schmitt}, \citenamefont {Lu}, \citenamefont {Moore},
  \citenamefont {Zhang}, \citenamefont {Meevasana}, \citenamefont {Sasagawa},
  \citenamefont {Uchida}, \citenamefont {Fujita}, \citenamefont {Ishida},
  \citenamefont {Ishikado}, \citenamefont {Yoshida}, \citenamefont {Eisaki},
  \citenamefont {Hussain}, \citenamefont {Devereaux},\ and\ \citenamefont
  {Shen}}]{Vishik2012}%
  \BibitemOpen
  \bibfield  {author} {\bibinfo {author} {\bibfnamefont {I.~M.}\ \bibnamefont
  {Vishik}}, \bibinfo {author} {\bibfnamefont {M.}~\bibnamefont {Hashimoto}},
  \bibinfo {author} {\bibfnamefont {R.-H.}\ \bibnamefont {He}}, \bibinfo
  {author} {\bibfnamefont {W.-S.}\ \bibnamefont {Lee}}, \bibinfo {author}
  {\bibfnamefont {F.}~\bibnamefont {Schmitt}}, \bibinfo {author} {\bibfnamefont
  {D.}~\bibnamefont {Lu}}, \bibinfo {author} {\bibfnamefont {R.~G.}\
  \bibnamefont {Moore}}, \bibinfo {author} {\bibfnamefont {C.}~\bibnamefont
  {Zhang}}, \bibinfo {author} {\bibfnamefont {W.}~\bibnamefont {Meevasana}},
  \bibinfo {author} {\bibfnamefont {T.}~\bibnamefont {Sasagawa}}, \bibinfo
  {author} {\bibfnamefont {S.}~\bibnamefont {Uchida}}, \bibinfo {author}
  {\bibfnamefont {K.}~\bibnamefont {Fujita}}, \bibinfo {author} {\bibfnamefont
  {S.}~\bibnamefont {Ishida}}, \bibinfo {author} {\bibfnamefont
  {M.}~\bibnamefont {Ishikado}}, \bibinfo {author} {\bibfnamefont
  {Y.}~\bibnamefont {Yoshida}}, \bibinfo {author} {\bibfnamefont
  {H.}~\bibnamefont {Eisaki}}, \bibinfo {author} {\bibfnamefont
  {Z.}~\bibnamefont {Hussain}}, \bibinfo {author} {\bibfnamefont {T.~P.}\
  \bibnamefont {Devereaux}}, \ and\ \bibinfo {author} {\bibfnamefont {Z.-X.}\
  \bibnamefont {Shen}},\ }\href {\doibase 10.1073/pnas.1209471109} {\bibfield
  {journal} {\bibinfo  {journal} {Proc. Natl. Acad. Sci.}\ }\textbf {\bibinfo
  {volume} {109}},\ \bibinfo {pages} {18332} (\bibinfo {year} {2012})},\
  \Eprint {http://arxiv.org/abs/1209.6514} {1209.6514} \BibitemShut {NoStop}%
\bibitem [{\citenamefont {Montiel}\ \emph
  {et~al.}(2017{\natexlab{b}})\citenamefont {Montiel}, \citenamefont {Kloss},\
  and\ \citenamefont {P{\'e}pin}}]{Montiel2017sr}%
  \BibitemOpen
  \bibfield  {author} {\bibinfo {author} {\bibfnamefont {X.}~\bibnamefont
  {Montiel}}, \bibinfo {author} {\bibfnamefont {T.}~\bibnamefont {Kloss}}, \
  and\ \bibinfo {author} {\bibfnamefont {C.}~\bibnamefont {P{\'e}pin}},\ }\href
  {\doibase 10.1038/s41598-017-01538-1} {\bibfield  {journal} {\bibinfo
  {journal} {Scientific Reports}\ }\textbf {\bibinfo {volume} {7}},\ \bibinfo
  {pages} {3477} (\bibinfo {year} {2017}{\natexlab{b}})}\BibitemShut {NoStop}%
\bibitem [{\citenamefont {Efetov}\ \emph {et~al.}(2013)\citenamefont {Efetov},
  \citenamefont {Meier},\ and\ \citenamefont {P\'epin}}]{Efetov13}%
  \BibitemOpen
  \bibfield  {author} {\bibinfo {author} {\bibfnamefont {K.~B.}\ \bibnamefont
  {Efetov}}, \bibinfo {author} {\bibfnamefont {H.}~\bibnamefont {Meier}}, \
  and\ \bibinfo {author} {\bibfnamefont {C.}~\bibnamefont {P\'epin}},\ }\href
  {\doibase 10.1038/nphys2641} {\bibfield  {journal} {\bibinfo  {journal} {Nat.
  Phys.}\ }\textbf {\bibinfo {volume} {9}},\ \bibinfo {pages} {442} (\bibinfo
  {year} {2013})}\BibitemShut {NoStop}%
\bibitem [{\citenamefont {Montiel}\ \emph
  {et~al.}(2016{\natexlab{a}})\citenamefont {Montiel}, \citenamefont {Kloss},
  \citenamefont {P\'epin}, \citenamefont {Benhabib}, \citenamefont {Gallais},\
  and\ \citenamefont {Sacuto}}]{Montiel15a}%
  \BibitemOpen
  \bibfield  {author} {\bibinfo {author} {\bibfnamefont {X.}~\bibnamefont
  {Montiel}}, \bibinfo {author} {\bibfnamefont {T.}~\bibnamefont {Kloss}},
  \bibinfo {author} {\bibfnamefont {C.}~\bibnamefont {P\'epin}}, \bibinfo
  {author} {\bibfnamefont {S.}~\bibnamefont {Benhabib}}, \bibinfo {author}
  {\bibfnamefont {Y.}~\bibnamefont {Gallais}}, \ and\ \bibinfo {author}
  {\bibfnamefont {A.}~\bibnamefont {Sacuto}},\ }\href {\doibase
  10.1103/PhysRevB.93.024515} {\bibfield  {journal} {\bibinfo  {journal} {Phys.
  Rev. B}\ }\textbf {\bibinfo {volume} {93}},\ \bibinfo {pages} {024515}
  (\bibinfo {year} {2016}{\natexlab{a}})}\BibitemShut {NoStop}%
\bibitem [{\citenamefont {Montiel}\ \emph
  {et~al.}(2016{\natexlab{b}})\citenamefont {Montiel}, \citenamefont {Kloss},\
  and\ \citenamefont {P{\'e}pin}}]{Montiel:2016it}%
  \BibitemOpen
  \bibfield  {author} {\bibinfo {author} {\bibfnamefont {X.}~\bibnamefont
  {Montiel}}, \bibinfo {author} {\bibfnamefont {T.}~\bibnamefont {Kloss}}, \
  and\ \bibinfo {author} {\bibfnamefont {C.}~\bibnamefont {P{\'e}pin}},\ }\href
  {\doibase 10.1209/0295-5075/115/57001} {\bibfield  {journal} {\bibinfo
  {journal} {EPL (Europhysics Letters)}\ }\textbf {\bibinfo {volume} {115}},\
  \bibinfo {pages} {57001} (\bibinfo {year} {2016}{\natexlab{b}})}\BibitemShut
  {NoStop}%
\bibitem [{\citenamefont {Montiel}\ and\ \citenamefont
  {P{\'{e}}pin}()}]{MontielNeutrons2017}%
  \BibitemOpen
  \bibfield  {author} {\bibinfo {author} {\bibfnamefont {X.}~\bibnamefont
  {Montiel}}\ and\ \bibinfo {author} {\bibfnamefont {C.}~\bibnamefont
  {P{\'{e}}pin}},\ }\href {http://arxiv.org/abs/1703.04442} {\ }\Eprint
  {http://arxiv.org/abs/1703.04442} {arXiv:1703.04442} \BibitemShut {NoStop}%
\bibitem [{\citenamefont {Hsu}\ \emph {et~al.}(2017)\citenamefont {Hsu},
  \citenamefont {Hartstein}, \citenamefont {Davies}, \citenamefont {Chan},
  \citenamefont {Porras}, \citenamefont {Loew}, \citenamefont {Taylor},
  \citenamefont {Liu}, \citenamefont {Tacon}, \citenamefont {Zuo},
  \citenamefont {Wang}, \citenamefont {Zhu}, \citenamefont {Lonzarich},
  \citenamefont {Keimer}, \citenamefont {Harrison},\ and\ \citenamefont
  {Sebastian}}]{Hsu2017}%
  \BibitemOpen
  \bibfield  {author} {\bibinfo {author} {\bibfnamefont {Y.-T.}\ \bibnamefont
  {Hsu}}, \bibinfo {author} {\bibfnamefont {M.}~\bibnamefont {Hartstein}},
  \bibinfo {author} {\bibfnamefont {A.~J.}\ \bibnamefont {Davies}}, \bibinfo
  {author} {\bibfnamefont {M.~K.}\ \bibnamefont {Chan}}, \bibinfo {author}
  {\bibfnamefont {J.}~\bibnamefont {Porras}}, \bibinfo {author} {\bibfnamefont
  {T.}~\bibnamefont {Loew}}, \bibinfo {author} {\bibfnamefont {S.~V.}\
  \bibnamefont {Taylor}}, \bibinfo {author} {\bibfnamefont {H.}~\bibnamefont
  {Liu}}, \bibinfo {author} {\bibfnamefont {M.~L.}\ \bibnamefont {Tacon}},
  \bibinfo {author} {\bibfnamefont {H.}~\bibnamefont {Zuo}}, \bibinfo {author}
  {\bibfnamefont {J.}~\bibnamefont {Wang}}, \bibinfo {author} {\bibfnamefont
  {Z.}~\bibnamefont {Zhu}}, \bibinfo {author} {\bibfnamefont {G.~G.}\
  \bibnamefont {Lonzarich}}, \bibinfo {author} {\bibfnamefont {B.}~\bibnamefont
  {Keimer}}, \bibinfo {author} {\bibfnamefont {N.}~\bibnamefont {Harrison}}, \
  and\ \bibinfo {author} {\bibfnamefont {S.~E.}\ \bibnamefont {Sebastian}},\
  }\href@noop {} {\bibfield  {journal} {\bibinfo  {journal} {in preparation}\ }
  (\bibinfo {year} {2017})}\BibitemShut {NoStop}%
\bibitem [{\citenamefont {Voruganti}\ \emph {et~al.}(1992)\citenamefont
  {Voruganti}, \citenamefont {Golubentsev},\ and\ \citenamefont
  {John}}]{Voruganti1992}%
  \BibitemOpen
  \bibfield  {author} {\bibinfo {author} {\bibfnamefont {P.}~\bibnamefont
  {Voruganti}}, \bibinfo {author} {\bibfnamefont {A.}~\bibnamefont
  {Golubentsev}}, \ and\ \bibinfo {author} {\bibfnamefont {S.}~\bibnamefont
  {John}},\ }\href@noop {} {\bibfield  {journal} {\bibinfo  {journal} {Phys.
  Rev. B}\ }\textbf {\bibinfo {volume} {45}},\ \bibinfo {pages} {13945}
  (\bibinfo {year} {1992})}\BibitemShut {NoStop}%
\bibitem [{\citenamefont {Eberlein}\ \emph {et~al.}(2016)\citenamefont
  {Eberlein}, \citenamefont {Metzner}, \citenamefont {Sachdev},\ and\
  \citenamefont {Yamase}}]{Eberlein2016}%
  \BibitemOpen
  \bibfield  {author} {\bibinfo {author} {\bibfnamefont {A.}~\bibnamefont
  {Eberlein}}, \bibinfo {author} {\bibfnamefont {W.}~\bibnamefont {Metzner}},
  \bibinfo {author} {\bibfnamefont {S.}~\bibnamefont {Sachdev}}, \ and\
  \bibinfo {author} {\bibfnamefont {H.}~\bibnamefont {Yamase}},\ }\href
  {\doibase 10.1103/PhysRevLett.117.187001} {\bibfield  {journal} {\bibinfo
  {journal} {Phys. Rev. Lett.}\ }\textbf {\bibinfo {volume} {117}},\ \bibinfo
  {pages} {187001} (\bibinfo {year} {2016})}\BibitemShut {NoStop}%
\bibitem [{\citenamefont {Mahan}(2000)}]{Mahan}%
  \BibitemOpen
  \bibfield  {author} {\bibinfo {author} {\bibfnamefont {G.~D.}\ \bibnamefont
  {Mahan}},\ }\href@noop {} {\emph {\bibinfo {title} {Many-Particle Physics}}}\
  (\bibinfo  {publisher} {Springer},\ \bibinfo {year} {2000})\BibitemShut
  {NoStop}%
\bibitem [{\citenamefont {Yang}\ \emph
  {et~al.}(2006{\natexlab{b}})\citenamefont {Yang}, \citenamefont {Rice},\ and\
  \citenamefont {Zhang}}]{Yang:2006eq}%
  \BibitemOpen
  \bibfield  {author} {\bibinfo {author} {\bibfnamefont {K.-Y.}\ \bibnamefont
  {Yang}}, \bibinfo {author} {\bibfnamefont {T.~M.}\ \bibnamefont {Rice}}, \
  and\ \bibinfo {author} {\bibfnamefont {F.-C.}\ \bibnamefont {Zhang}},\ }\href
  {\doibase 10.1103/PhysRevB.73.174501} {\bibfield  {journal} {\bibinfo
  {journal} {Physical Review B}\ }\textbf {\bibinfo {volume} {73}} (\bibinfo
  {year} {2006}{\natexlab{b}}),\ 10.1103/PhysRevB.73.174501}\BibitemShut
  {NoStop}%
\end{thebibliography}%

\widetext
\clearpage
\begin{center}
\textbf{\large Supplemental material for: Evolution of Hall resistivity and spectral function with doping in the SU(2) theory of cuprates}
\end{center}
\setcounter{equation}{0}
\setcounter{figure}{0}
\setcounter{table}{0}
\setcounter{page}{1}
\setcounter{section}{0}
\makeatletter
\renewcommand{\theequation}{S\arabic{equation}}
\renewcommand{\thefigure}{S\arabic{figure}}
\renewcommand{\bibnumfmt}[1]{[S#1]}

\section{Derivation of the self-energy}

Here we give the details of the derivation of the self-energy described in the main text (Eq.\ 1 to 5). Previous work (\cite{Montiel2017} Eq.\ 83) gives us the partition function for the SU(2) fluctuations:
\begin{equation}
Z_{fin}=e^{-S_{fin}}=e^{\frac{1}{2} \left\langle S_{int}^2 \right\rangle_Q}
\end{equation}
where the average is over the effective SU(2) fluctuation modes $Q$. We approximate $S_{fin}$ to its effective component in the superconducting channel (\cite{Montiel2017} Eq.\ 86):
\begin{equation}
S_{fin} = -\frac{1}{2} \Tr \sum_{k,k',q,q',\sigma,\sigma'} \sigma \sigma' \left\langle \Delta_{kq}^\dagger \Delta_{k'q'} \right\rangle_Q \psi^\dagger_{k+q,\sigma} \psi^\dagger_{-k+q,\bar{\sigma}} \psi_{-k'+q',\bar{\sigma}'} \psi_{k'+q',\sigma'}
\end{equation}
This is justified by the fact that the rest of the action, namely its effective component in the charge channel, has been shown to stabilise pair density wave orders \cite{Montiel2017}. These are second order and can safely be neglected for the calculation of the SU(2) fluctuations \cite{Montiel2017}. The non-linear $\sigma$-model yields (\cite{Montiel2017} Eq.\ 88):
\begin{equation}
\left\langle \Delta_{kq}^\dagger \Delta_{k'q'} \right\rangle_Q = \delta_{\textbf{q},\textbf{q}'} \pi^s_{kk'q}
\end{equation}
where $\pi^s_{kk'q}$ is the SU(2) propagator (\cite{Montiel2017} Eq.\ 91):
\begin{equation}
\pi^s_{kk'q} = M_{0,k} M_{0,k'} \frac{\pi_0}{J_0 \epsilon^2 + J_1(\textbf{v }\cdot \textbf{q})^2 - a_0}
\end{equation}
where $a_0$ is the mass in the non-linear $\sigma$-model, $\textbf{v}$ is the Fermi velocity and $J_0$ and $J_1$ are given by the non-linear $\sigma$-model. Superconductivity is favoured if $a_0>0$, and the charge order is favoured if $a_0<0$, these two orders still being linked by the SU(2) constraint. In particular, $a_0$ varies with magnetic field: it becomes negative when one is applied \cite{Montiel2017}. We can input this in our expression for the action:
\begin{align}
S_{fin} = &-\frac{1}{2} \sum_{k,k',q,\sigma,\sigma'} \sigma \sigma' \pi^s_{kk'q} \Tr \psi^\dagger_{k+q,\sigma} \psi^\dagger_{-k+q,\bar{\sigma}} \psi_{-k'+q,\bar{\sigma}'} \psi_{k'+q,\sigma'}
\\
S_{fin} = &-\frac{1}{2} \sum_{k,q,\sigma} \pi^s_{kkq} \psi^\dagger_{k+q,\sigma} G^0_{-k+q,\bar{\sigma}} \psi_{k+q,\sigma}
\end{align}
where $G^0$ is the bare electron propagator. We obtain the self-energy:
\begin{align}
-\Sigma(k) = &\frac{1}{2} \sum_{q,\sigma} \pi^s_{kk'q} G^0_{-k+q,\bar{\sigma}}
\\
\Sigma(k) = &-\frac{1}{2} \sum_{q,\sigma} \frac{M_{0,k}^2 \pi_0}{J_0 \epsilon^2 + J_1(\textbf{v }\cdot \textbf{q})^2 - a_0} \frac{1}{i(-\omega+\epsilon) - \xi_{-k+q}}
\\
= & M_{0,k}^2 \pi_0 \sum_{q} \frac{1}{J_0 \epsilon^2 + J_1(\textbf{v}\cdot \textbf{q})^2 - a_0} \frac{1}{i\omega - i\epsilon + \xi_{k-q}}
\end{align}
We now can name $q_{\parallel}$ the magnitude of the component of $\textbf{q}$ parallel to $\textbf{v}$. Approximating $\xi_{k-q}$ by its component in $\textbf{q}$ close to the Fermi surface gives:
\begin{equation}
\Sigma(k) = M_{0,k}^2 \pi_0 \sum_{\omega}  \int_{-\infty}^{\infty} dq_\parallel \frac{1}{J_0 \epsilon^2 + J_1 v^2 q_\parallel^2 - a_0} \frac{1}{i\omega - i\epsilon + \xi_{k} - v q_\parallel}
\end{equation}
where $v = |\textbf{v}|$. Close to the Fermi surface, it has been shown that $J_0=J_1$ hence we can set them both to one without loss of generality \cite{Montiel2017}. Finally, we neglect the mass $a_0$ and obtain:
\begin{equation}
\Sigma(k) = M_{0,k}^2 \pi_0 \sum_{\omega} \int_{-\infty}^{\infty} dq_\parallel \frac{1}{\epsilon^2 + v^2 q_\parallel^2} \frac{1}{i\omega - i\epsilon + \xi_{k} - v q_\parallel}
\end{equation}
In order to integrate on $vq_\parallel$, we define the function in the integral on the whole complex plane and integrate this function over a contour surrounding one half of the plane, given that this function goes to zero when $|vq_\parallel| \to \infty$. We choose the half plane which contains the pole $-i \epsilon$. If $(\omega - \epsilon) \times sgn(\epsilon) < 0$, it contains a second pole: $i\omega - i\epsilon + \xi_{k}$ . The residues theorem gives:
\begin{align}
\int_{-\infty}^{\infty} dvq_\parallel \frac{1}{\epsilon^2 + v^2 q_\parallel^2} \frac{1}{i\omega - i\epsilon + \xi_{k} - v q_\parallel} = &\frac{\pi}{\omega} \times \frac{1}{i\omega + \xi_{k}} + \frac{2\pi i \Theta[(\epsilon - \omega)sgn(\epsilon)]}{\epsilon^2 + (i\omega - i\epsilon + \xi_{k})^2}
\\
= &\frac{\pi}{\epsilon} \times \frac{1 - \Theta[(\epsilon - \omega)sgn(\epsilon)]}{i\omega + \xi_{k}} + \frac{\pi}{\epsilon} \times \frac{ \Theta[(\epsilon - \omega)sgn(\epsilon)]}{i\omega - 2i\epsilon + \xi_{k}}
\end{align}
where $\Theta$ is the step function. Our expression for the self-energy then becomes:
\begin{align}
\Sigma(k) = & \frac{M_{0,k}^2}{i\omega + \xi_{k}} \frac{\pi_0 \pi}{v} \sum_{\epsilon} \frac{1 - \Theta[(\epsilon - \omega)sgn(\epsilon)]}{\epsilon} + \frac{M_{0,k}^2 \pi_0 \pi}{v} \sum_{\epsilon} \frac{\Theta[(\epsilon - \omega)sgn(\epsilon)]}{\epsilon \left( i\omega - 2i\epsilon + \xi_{k} \right)}
\end{align}
We now approximate this sum by its term linear in $1/\epsilon$ and sum over $\epsilon$. Note that we need to reintroduce the small mass $\sqrt{-a_0}$ for the sum to converge. We thus obtain the final expression for the self-energy:
\begin{equation}
\Sigma(k) = B \frac{M_{0,k}^2}{i\omega + \xi_{k}}
\end{equation}
where $B$ is the prefactor including the sum on $\epsilon$ and the other prefactors.

\section{Conductivities}

We calculated the longitudinal and transverse conductivities in order to calculate the Hall resistance. For completeness, here we plot them as a function of temperature for a range of hole dopings.

\begin{figure}[h]
\centering
\includegraphics[width=8cm]{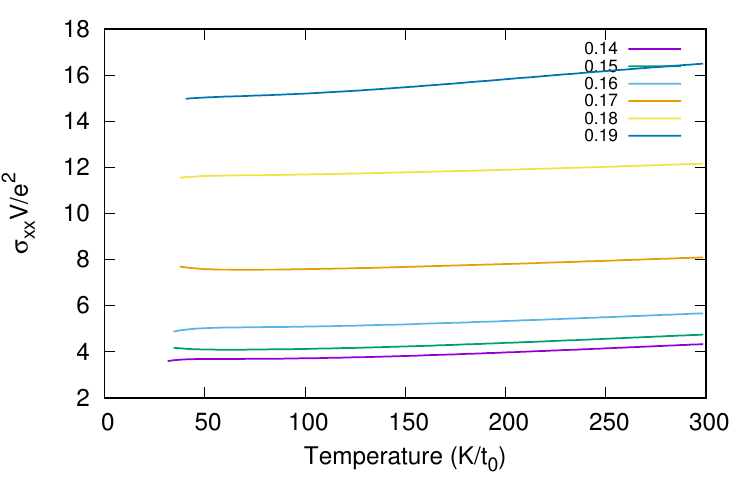}
\caption{Longitudinal conductivity with respect to temperature for a range of hole dopings.}
\label{fig:conductivityXX}
\end{figure}

\begin{figure}[h]
\centering
\includegraphics[width=8cm]{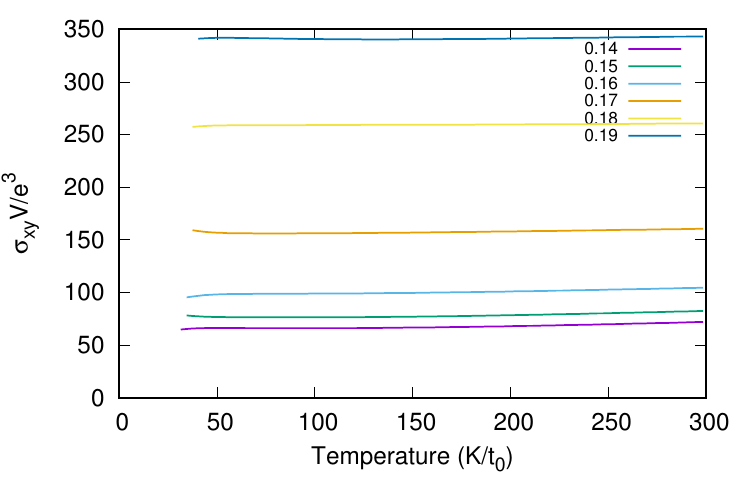}
\caption{Transverse conductivity with respect to temperature for a range of hole dopings.}
\label{fig:conductivityXY}
\end{figure}

\section{$d$-wave gap}

\begin{figure}[h]
\centering
\includegraphics[width=8cm]{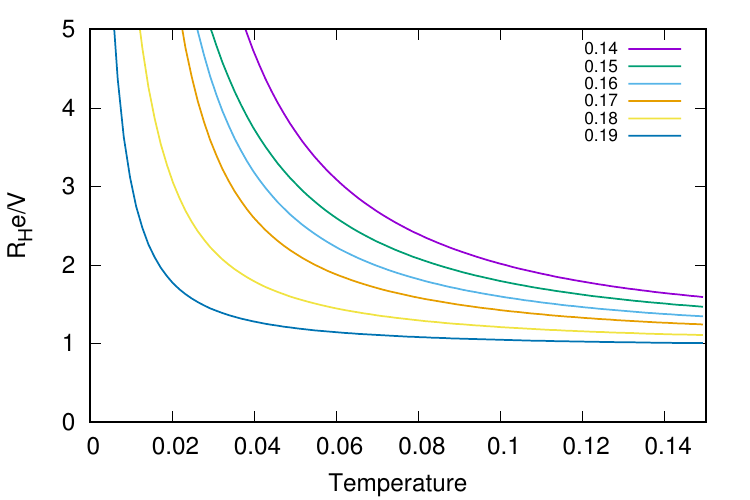}
\caption{Hall resistance per volume unit with respect to temperature in units of $t_0 / k_B$ for a range of hole dopings.}
\label{fig:RH-d-wave}
\end{figure}

We compared the results we obtained using the SU(2) gap based on the SU(2) symmetry-breaking term to a standard $d$-wave gap used in previous studies \cite{Yang:2006eq,Storey2016}
\begin{equation}
B M_{0,k}^2 = \left[ \frac{3 t_0}{2} (0.2-x) \left (cos(k_x) - cos(k_y) \right) \right]^2
\end{equation}
Because the dispersion of the bosonic mode is the opposite of the electronic dispersion, the gap opens everywhere but at the nodal point. Consequently, the Hall resistivity diverges at low temperature (Figure \ref{fig:RH-d-wave})

\begin{figure}
\centering
\includegraphics[width=8cm]{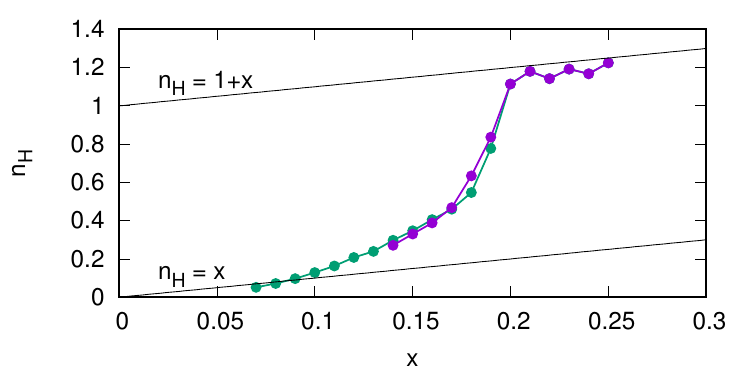}
\caption{Hall number at $k_B T = 0.05 t_0 / k_B$ for a $d$-wave gap with respect to doping (purple), and Hall number at zero temperature in the Yang-Rice-Zhang model (green). The low-doping and high-doping asymptotes are also plotted for reference.}
\label{fig:nH-d-wave}
\end{figure}

One can however study the evolution of the Hall number at finite temperature. Interestingly, this evolution reproduces closely the one of the Hall number extrapolated to zero temperature in the Yang-Rice-Zhang model (Figure \ref{fig:nH-d-wave}).

\section{Other choices of ordering wave vector}

Following previous work on the SU(2) theory of cuprate superconductors \cite{Montiel2017}, we chose $Q_0$ as the vector between the two closest hot spots in the main text. Here we explore the consequences of making a different choice. We used two different ordering vectors: the vector linking two points of the Fermi surface on the Brillouin zone-edge, and a diagonal vector linking two hot spots.

We adjusted the gap $\Delta_{SU2}$ in magnitude for the zone-edge $Q_0$ vector:
\begin{equation}
\Delta_{SU2} = \left( \frac{1}{e^{(x - 0.175) \times 170} + 1} - 0.018 \right) \times 0.61
\end{equation}
and for the diagonal $Q_0$ vector
\begin{equation}
\Delta_{SU2} = \left( \frac{1}{e^{(x - 0.175) \times 170} + 1} - 0.018 \right) \times 0.22
\end{equation}
Note that the only change with the gap used in the main text is the prefactor. This change is very small for the zone-edge calculation: from 0.58 to 0.61, but much larger for the diagonal $Q_0$ vector where we use a prefactor of 0.22. A higher value of the prefactor results in $n_H$ dropping to zero at finite doping.

\begin{figure}
\centering
\includegraphics[width=8cm]{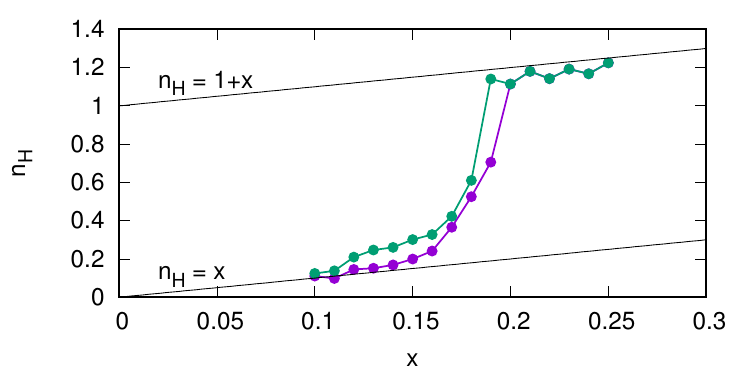}
\caption{Hall number with respect to doping for a zone-edge $Q_0$ (purple), and diagonal $Q_0$ (green). The low-doping and high-doping asymptotes are also plotted for reference.}
\label{fig:nH-otherQ}
\end{figure}

The evolution of the Hall number with doping is qualitatively similar in both cases with what is found in the main text (Figure \ref{fig:nH-otherQ}), except for the case of the diagonal $Q_0$ vector very close to the critical doping. Indeed, the $1+x$ asymptote is reached at $x=0.19$ already, due to the earlier closing of the gap close to the zone-edge. This is reminiscent of experiments which found similarly high Hall numbers before the closing of the pseudogap \cite{Collignon2016}.

\section{Closing of the gap close to the zone-edge}

\begin{figure}[h]
\centering
\includegraphics[width=8cm]{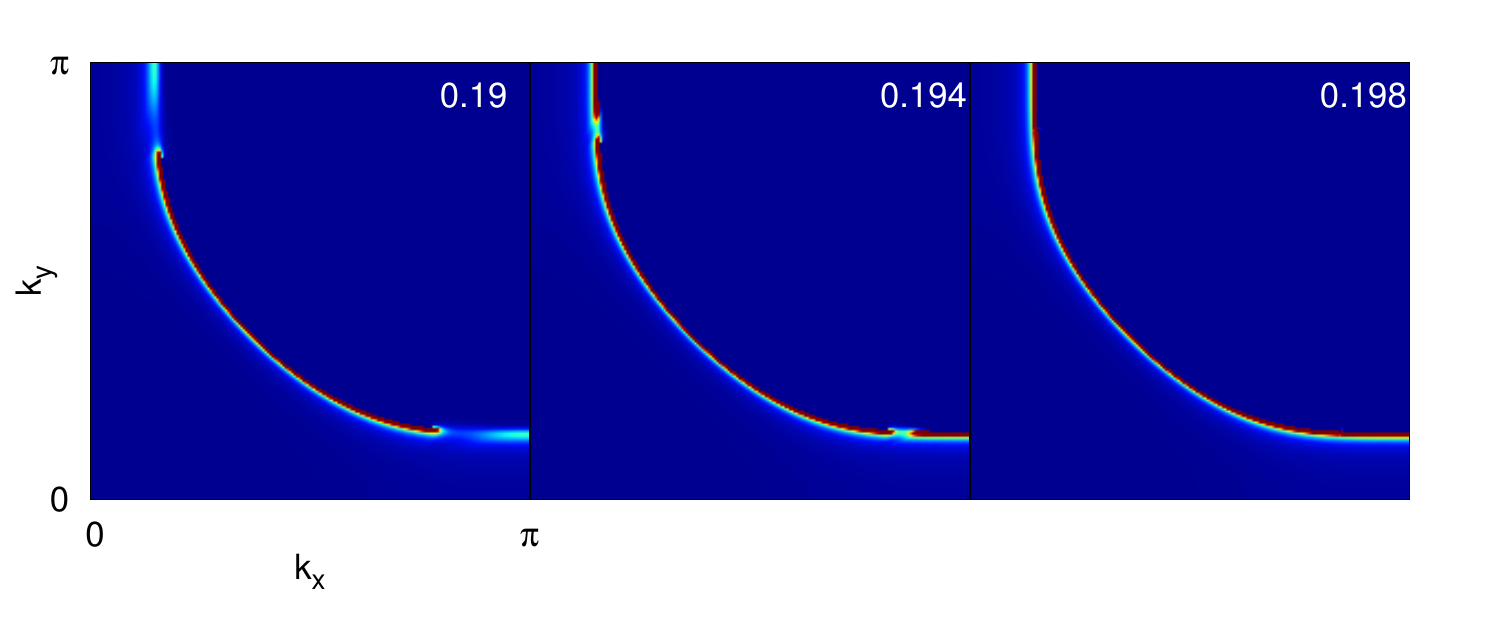}
\caption{Spectral function at high doping. The arising of the second Fermi arc crossing the Brillouin zone edge is clearly visible. Note that because the renormalised bands are symmetrical with respect to zero energy, both spectral functions are equal at zero frequency. We therefore only plot one of them.}
\label{fig:spectral-high-doping}
\end{figure}

Here we go back to the discussion on the case discussed in the main text, meaning longitudinal ordering wave vectors linking hot spots. As discussed in the main text, the Fermi arc opens close to the nodal point and widens when the hole doping increases. Given that the gap opens at the hot spots, it closes near the Brillouin zone edge before the transition. A second Fermi arc therefore appears, crossing the zone edge (Figure \ref{fig:spectral-high-doping}).

\begin{figure}
\centering
\includegraphics[width=8cm]{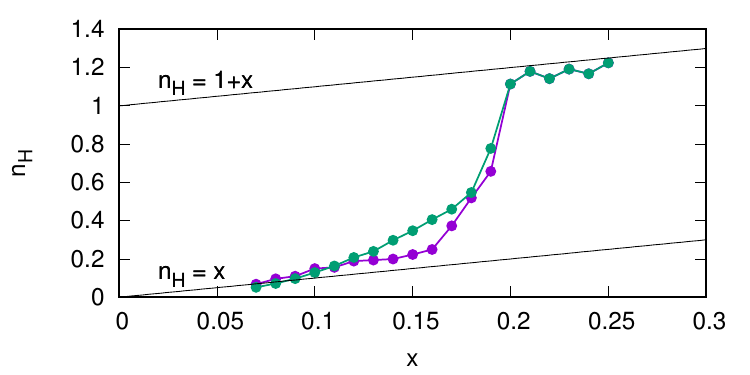}
\caption{Hall number with respect to doping in the SU(2) theory (purple), and Yand-Rice-Zhang (YRZ) model (green). The low-doping and high-doping asymptotes are also plotted for reference. Note that the data for the YRZ model crosses the asymptote at low doping.}
\label{fig:nH-Storey}
\end{figure}

\section{Comparison with the YRZ model}

We compared the doping dependence of the Hall number obtained within the SU(2) theory to the one obtained using the Yang-Rice-Zhang (YRZ) model. The dependence is similar, except that the transition is slightly sharper in our case (Figure \ref{fig:nH-Storey}).

We also note that the Hall number at low doping in the YRZ model is below the $n_H = x$ line. This is due to a value of the gap which is too large. We therefore believe that describing the pseudogap by a linear function in doping is inadequate. This does not however hinder the validity of the YRZ model.

\end{document}